\documentclass{aa}  

\usepackage{enumitem}

\usepackage{orcidlink}

\usepackage{xspace}
\newcommand{\Ha}{\ifmmode {\mathrm{H}\alpha} \else H$\alpha$\fi\xspace}
\newcommand{\Hb}{\ifmmode {\mathrm{H}\beta} \else H$\beta$\fi\xspace}
\newcommand{\hii}{H$\,${\sc ii}\xspace}

\newcommand{\Hii}{\ifmmode \rm{H}\,\textsc{ii} \else H~{\textsc{ii}}\fi\xspace}
\newcommand{\Hi}{\ifmmode \rm{H}\,\textsc{i} \else H~{\textsc{i}}\fi\xspace}
\newcommand{\Nii}{\ifmmode [\text{N}\,\textsc{ii}]\lambda 6584 \else [N~{\scshape ii}]$\lambda 6584$\fi\xspace}
\newcommand{\nii}{\ifmmode [\text{N}\,\textsc{ii}] \else [N~{\scshape ii}]\fi\xspace}
\newcommand{\Oii}{\ifmmode [\rm{O}\,\textsc{ii}]\lambda 3727 \else [O~{\textsc{ii}}]$\lambda$3727\fi}
\newcommand{\oii}{\ifmmode [\rm{O}\,\textsc{ii}] \else [O~{\textsc{ii}}]\fi}
\newcommand{\Oiii}{\ifmmode [\rm{O}\,\textsc{iii}]\lambda 5007 \else [O~{\textsc{iii}}]$\lambda$5007\fi}
\newcommand{\oiii}{\ifmmode [\rm{O}\,\textsc{iii}] \else [O~{\textsc{iii}}]\fi}

\usepackage{graphicx}
\usepackage{txfonts}

\begin{document}

   \title{Gas Metallicity of Ram-Pressure Stripped Galaxies at Intermediate Redshift with MUSE Data}

%   \subtitle{I. Spatially-Resolved Analysis of RPS galaxies in Clusters Abell370 and Abell2744}

   \author{A. Khoram 
          \inst{1,2}
          \orcidlink{https://orcid.org/0009-0009-6563-282X}
          \and
          B. Poggianti\inst{2}
          \orcidlink{https://orcid.org/0000-0001-8751-8360}
          \and
          A. Moretti\inst{2}
          \orcidlink{https://orcid.org/0000-0002-1688-482X}
           \and
          B. Vulcani\inst{2}
          \orcidlink{https://orcid.org/0000-0003-0980-1499}
          \and
          M. Radovich\inst{2}
          \orcidlink{https://orcid.org/0000-0002-3585-866X}
          \and
          A. Werle\inst{2}
          \orcidlink{https://orcid.org/0000-0002-4382-8081}
          \and
          M. Gullieuszik\inst{2}
          \orcidlink{https://orcid.org/0000-0002-7296-9780}
          \and
          J. Richard\inst{3}
          \orcidlink{https://orcid.org/0000-0001-5492-1049}
          %\fnmsep\thanks{}
}

   \institute{Dipartimento di Fisica e Astronomia “G. Galilei”, Universit\`a degli Studi di Padova, via Marzolo 8, I-35131 Padova, Italy
              %\email{amirhossein.khoram@inaf.it}
         \and
            INAF-Osservatorio Astronomico di Padova, vicolo dell’Osservatorio 5, 35122 Padova, Italy
         \and 
         Univ. Lyon, Univ Lyon1, ENS de Lyon, CNRS, Centre de Recherche Astrophysique de Lyon UMR5574, F-69230 Saint-Genis-Laval, France
             %\thanks{}
             }

   \date{}

  \abstract{
  Extraplanar tails of ionized stripped gas, extending up to several tens of kiloparsecs beyond the stellar disk, are often observed in ram-pressure stripped (RPS) galaxies in low redshift clusters. Recent studies have identified similar tails also at high redshift and we here present the first analysis of the chemical composition of such tails beyond the local universe. 
Specifically, we examine the distribution of ionized gas metallicity of RPS galaxies in the Abell 2744 (z=0.308) and Abell 370 (z=0.375) clusters observed as part of the MUSE-GTO program. We investigate spatially-resolved and global metallicities in galactic disks and stripped tails, utilizing 
both a theoretical calibration through a photoionization model and an empirical calibration. 
The metallicity gradients and the spatially resolved mass-metallicity relations indicate that the metallicity in the tails reaches up to $\sim 0.6$dex lower values than anywhere in the parent disks, with a few exceptions.
%{\bf for low galaxy stellar masses?}\ak{but we also have A2744-09!}
Both disks and tails follow a global mass-metallicity relation, though the tail metallicity is systematically lower than the one of the corresponding disk by up to $\sim 0.2$ dex. These findings demonstrate that additional processes are at play in the tails, and are consistent with a scenario of progressive dilution of metallicity along the tails due to the mixing of intracluster medium and interstellar gas, in accord with previous low-z results. In principle, the same scenario can also explain the flat or positive metallicity gradients observed in low-mass RPS galaxies, as in these galaxies the interstellar medium's metallicity can approach the metallicity levels found in the intracluster medium. 
}

  \keywords{}

  \maketitle
%
%-------------------------------------------------------------------

\section{Introduction}\label{sec:Intro}

The hot intracluster medium (ICM) can exert ram pressure on the interstellar medium (ISM) \citep{1972ApJ...176....1G} of galaxies within massive dark matter haloes, effectively stripping gas from them. This process is regarded as the most significant physical factor shaping the gas properties of cluster galaxies \citep{2006PASP..118..517B}. Observations have demonstrated numerous examples of ram-pressure stripped (RPS) galaxies in clusters, including the most extreme cases, dubbed jellyfish galaxies (e.g., \citealt{2014MNRAS.445.4335F} and \citealt{2016AJ....151...78P}).

Several observational studies in the local universe have investigated characteristics of the gas that has been stripped from galaxies (e.g., \citealt{2014A&A...570A..69B}; \citealt{2017ApJ...839..114J};   \citealt{2019MNRAS.482.4466P}; \citealt{2021ApJ...922..131T}; and \citealt{2023ApJ...950...24B}). They revealed a diverse gas composition, from radio emission \citep{2023arXiv230519941I}, neutral hydrogen \citep{2009AJ....138.1741C} and molecular gas \citep{2020ApJ...897L..30M} to X-ray emitting gas \citep{2010ApJ...708..946S}. The majority of these galaxies exhibit elongated and distorted tails of ionized gas that have been stripped up to several tens of kpc from the galaxy as it moves through the ICM (e.g. \citealt{2019ApJ...887..155P}). The extent of the stripping can vary depending on the properties of the galaxy, such as its velocity, mass, and inclination with respect to the ICM, as well as the density and pressure of the ICM \citep{2020ApJ...899...13G}.

Understanding the properties and behavior of the ionized gas in RPS galaxies is essential for unraveling the complex interplay between galaxies and their environments. One key aspect essential for comprehending the nature of gas within galaxies is the gas-phase metallicity (i.e., 12+log(O/H)). The gas-phase metallicity within the disk and along the stripped gas tail of such galaxies provides valuable insights into the interaction between the ISM and ICM during ram-pressure stripping. To date, the investigation of chemical compositions in RPS galaxies has been chiefly constrained to a handful of studies (\citealt{2016MNRAS.455.2028F}; \citealt{2017ApJ...846...27G};  \citealt{2018MNRAS.475.4055M}; \citealt{2019MNRAS.485.1157B}; \citealt{2020ApJ...895..106F}; \citealt{2021ApJ...923...28F}) conducted exclusively within the confines of the local universe ($z<0.1$). For example, within the GAs Stripping Phenomena in galaxies with MUSE (GASP, \citealt{2017ApJ...844...48P}) project, \cite{2021ApJ...922L...6F} used the MUSE emission lines data and a grid of photoionization models to derive the gas metallicity of 3 massive ($M_\star > 10^{10.95}M\odot$) RPS galaxies at redshift $\sim0.05$. The analyzed galaxy sample exhibited a steady reduction in metallicity within the tails as the distance from the galactic center increased. These findings led the authors to propose that the mixing of ISM and ICM is an essential process, corroborating the predictions made by simulations \citep{2021ApJ...911...68T}.

The gas-phase metallicity is also considered a good tracer of the global galaxy properties. In fact, a significant area of research has been dedicated to gaining insights into the correlation between global metallicity and other characteristics of galaxies \citep[e.g.][]{2004ApJ...613..898T, 2007ApJS..173..441H, 2010MNRAS.408.2115M, 2013A&A...550A.115H}. Considerable attention has been given to the mass-metallicity relation (MZR), the strong correlation between galaxy stellar mass and gas-phase oxygen abundance, by analyzing samples of star-forming galaxies at different redshifts (ranging from $0$ to $\sim 10$). The MZR of galaxies serves as a key empirical constraint for models of galactic evolution that try to explain how galaxies form throughout cosmic time. Noteworthy studies \citep[e.g.][]{2004ApJ...613..898T,2008ApJ...681.1183K,2008A&A...488..463M,2014ApJ...791..130Z,2015ApJ...799..138S,2020Jones,2023Nakajima} have contributed to this body of knowledge. Conversely, to date, only one study \citep{2020ApJ...895..106F} primarily focused on RPS galaxies and attempted to investigate their MZR at z$\sim$0. Likewise, the relationship between the stellar-mass surface density $\Sigma_{\star}$ and its local metallicity, known as the spatially-resolved mass-metallicity relation (rMZR), has been extensively explored over the past decade primarily in galaxies not subject to ram-pressure stripping within the local Universe (e.g., \citealt{2012ApJ...756L..31R}; \citealt{2013A&A...554A..58S}; \citealt{2018ApJ...868...89G}), yet investigations focusing on galaxies experiencing RPS remain notably absent. Due to its ability to accurately recreate the global metallicity and the metallicity profile over the galactocentric radius, the rMZR is seen by some authors to be a more fundamental relationship than the global MZR \citep{2016MNRAS.463.2513B}.\\

The ongoing pursuit is to discern whether the impact of ram-pressure stripping varies over cosmic time. Specifically, The primary objective is to examine the interaction between the ISM and the ICM during ram-pressure stripping through the window of analyzing their chemical compositions in clusters at z$\sim0.35$. This paper exploits the MUSE Guaranteed Time Observations (GTO) from \cite{2021A&A...646A..83R} to trace the gas-phase metallicity within the disks and along the tails of a sample of RPS galaxies spanning a wide stellar-mass range. We focus on the global, spatially-resolved, and radial distribution of ionized gas metallicity of Abell2744 and Abell370 RPS members \citep{2022ApJ...925....4M} with MUSE data.

The paper is organized as follows: In Sec. \ref{sec:galaxy sample}, we present the galaxy sample obtained from the two closest clusters observed with the MUSE GTO program. Sec. \ref{sec: data analysis} outlines the tools and methods used in our analysis. In Sec. \ref{sec: metallicity measurements}, we describe the methodologies employed to assess the gas-phase metallicity with indirect indicators. Sec. \ref{sec: results} presents this paper's main results, which are discussed in Sec. \ref{sec: Discussion}. Finally, Section  \ref{sec: summary} summarizes the conclusions of this study. 

\section{Galaxy and Cluster Sample}\label{sec:galaxy sample} 

%MUSE is a state-of-the-art integral-field spectrograph that utilizes an image-slicer with great efficacy. The MUSE deep data encompasses a wide range of comprehensive studies, offering a remarkable opportunity to delve into the intricate details of numerous clusters at intermediate redshifts, thanks to the MUSE GTO program \citep{2021A&A...646A..83R} which is also complemented by Hubble Deep and Ultra Deep Field data.\\

 \begin{figure}[ht]
    \centering
    \includegraphics[width=0.99\linewidth]{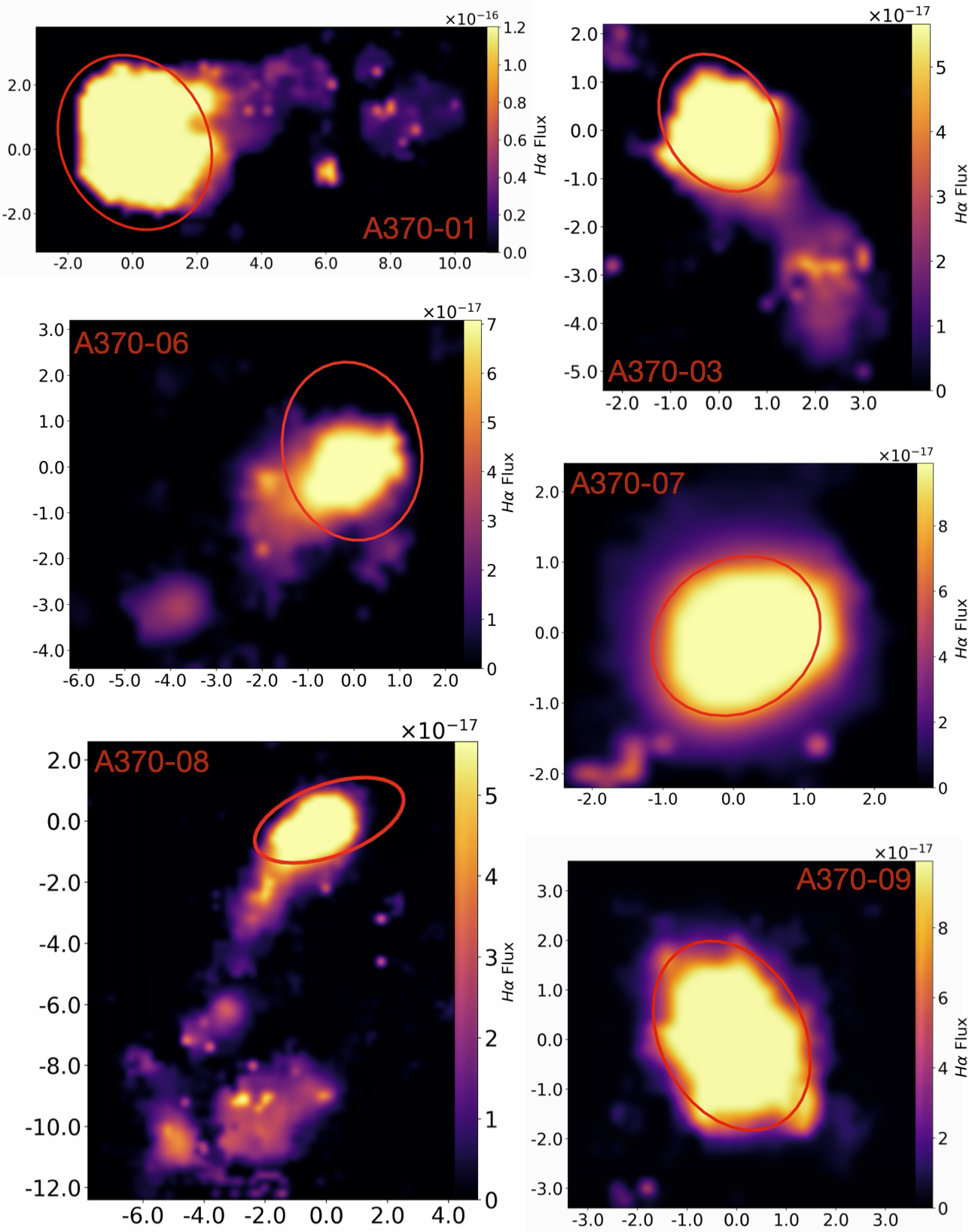}
    \caption{H\(\alpha\) [$\mathrm{erg\ 
    cm^{-2}s^{-1}arcsec^{-2}}$] maps of RPS galaxies in the Abell370 cluster obtained from MUSE data and smoothed with the bicubic interpolation method. The disk boundaries (sec \ref{disk_tail configuration}) are illustrated with red ellipses. Both axes are in the arcsec unit.}
    \label{fig:A370_moretti}
\end{figure}

\begin{figure}[ht]
    \centering
    \includegraphics[width=0.99\linewidth]{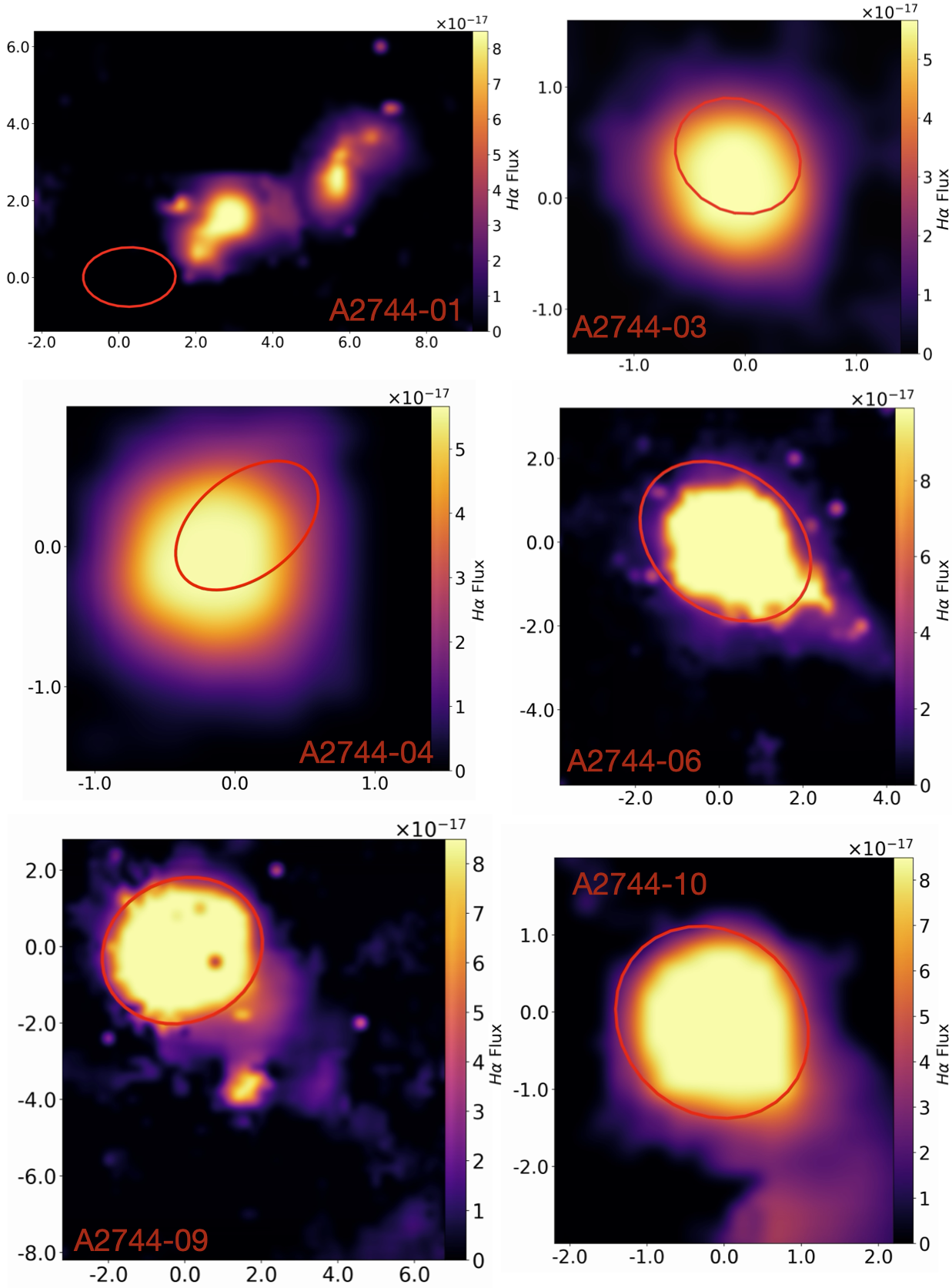}
    \caption{H\(\alpha\) [$\mathrm{erg\ 
    cm^{-2}s^{-1}arcsec^{-2}}$] maps of RPS galaxies in the Abell2744 cluster obtained from MUSE data and smoothed with the bicubic interpolation method. Both axes are in the arcsec unit and disk boundaries are shown with red ellipses. }
    \label{fig:2744_moretti}
\end{figure}

 In this paper, we work on Abell 370 and Abell 2744 (A370 and A2744, hereafter), the two most nearby clusters with the longest exposure time that have been observed within the MUSE GTO program \citep{2021A&A...646A..83R}. The clusters are part of the Frontier Fields (FFs) \citep{2017ApJ...837...97L}. Additionally, A2744 is also part of the MAssive Clusters Survey (MACS, \citealt{2001ApJ...553..668E}) and has also been the target of multiple JWST campaigns (e.g. \citealt{2022Treu} and \citealt{2022Bezanson}). For both clusters a \(2\times2\) MUSE mosaic exists, for a total exposure time of $20$hrs for A2744 and \(\sim 19\)hrs for A370 \citep{2021A&A...646A..83R} with excellent seeing ( $\sim 0.61-0.66$ arcsec). \\

We use the sample of galaxies classified as RPS by \cite{2022ApJ...925....4M}.  The authors examined MUSE data cubes and HST RGB images (F435W+F606W+F814W) to classify galaxies undergoing ram-pressure stripping. These were identified based on specific criteria: the presence of extraplanar, unilateral tails or debris with emission lines in the MUSE data cubes, and unilateral tails or debris visible in HST images, which were confirmed to be associated with the galaxy using MUSE redshift information. We have a total of 13 RPS galaxies at our disposal (Fig.\ref{fig:A370_moretti} and \ref{fig:2744_moretti}) which are members of the two clusters. Most of them have a stellar mass exceeding $10^9M_\odot$, with only three exceptions. Notably, one galaxy, A370-02, is excluded from the analysis due to an insufficient number of spaxels with well-detected H$\alpha$ emission lines. Additionally, in the case of A2744-01, nebular emission is exclusively present in the tail (Fig. \ref{fig:2744_moretti}) since the whole disk is in a post-starburst phase (lacking ionized gas, see \citealt{2022ApJ...930...43W}), our analysis exclusively focuses on the global properties of the tail in this galaxy.\\

A370 is a massive cluster at redshift \(0.375\) \citep{2009MNRAS.399.1447L} with the reported mass of \(8 \times 10^{14} M_{\odot}\) within \(500 \, \rm kpc\) and a velocity dispersion of \(\sim 1300 \rm \, km \, s^{-1}\) \citep{2017MNRAS.469.3946L}. This cluster contains $153$ spectroscopically confirmed members in the field covered by the MUSE mosaic. 
According to \cite{2022ApJ...925....4M}, after excluding A370-02 for the reasons mentioned earlier, the A370 cluster hosts 6 RPS galaxies (see Fig. \ref{fig:A370_moretti}) that are located from the center out to \( \sim 0.15R_{200} \) where \(R_{200} = 2.57 \, \rm Mpc\) \citep{2009MNRAS.399.1447L}. Moreover, \cite{2022ApJ...937...18B} demonstrated that although A370 is a merging cluster, there is no evidence to support the notion that the observed RPS galaxies are a consequence of the merger event. Rather, it is more probable that these galaxies are isolated infalling objects.\\

A2744 (also known as the Pandora cluster) is another intermediate redshift cluster at z=\(0.306\) \citep{2011ApJ...728...27O}. With a mass of \(\sim 2 \times 10^{15} M_{\odot}\) \citep{2016MNRAS.463.3876J}, it contains 227 spectroscopically confirmed members in the field covered by the MUSE mosaic. 
In the study conducted by \cite{2011ApJ...728...27O}, a comprehensive examination of the complex merging history of A2744 was carried out using X-ray and optical spectroscopy. They successfully identified two prominent substructures within the cluster, namely the northern core and the southern minor remnant core, along with a distinct region referred to as the central tidal debris. Remarkably, \cite{2022ApJ...937...18B} demonstrated that the 6 RPS galaxies in A2744 are part of a cluster substructure and that their high-speed encounter with the dense X-ray gas associated with the southern minor remnant core is likely responsible for the stripping.
%collision of the galaxies in the central tidal debris 
%with the X-ray gas associated with the southern minor remnant core is likely responsible for the stripping being experienced by the 6 RPS galaxies.

At the redshifts of these clusters, all main emission lines from $\lambda \sim3500$\AA \, up to $\sim 6800$\AA \, rest-frame are included in the MUSE spectra, thus we can derive dust correction and metallicity measurements with standard methods, as described below.
%(see \citealt{2019A&ARv..27....3M} for a review).

\section{Data Analysis}\label{sec: data analysis}

A number of spatially-resolved characteristics and parameters, as well as integrated quantities of the gaseous and stellar components of our galaxy sample can be analyzed through the study of MUSE data. Stellar mass, stellar disk definition, gas kinematics, and ionization mechanism characteristics of the sample have been already presented in the previous study, \cite{2022ApJ...925....4M}. For our analysis, we need to analyze disk and tail spectra and obtain the global and spatially-resolved chemical composition of the galaxy sample. In this section, we introduce the relevant tools and techniques.

\subsection{Disk/Tail Configuration}\label{disk_tail configuration}

We utilize the disk masks provided in \cite{2022ApJ...925....4M} based on the MUSE g-band reconstructed image to define the galaxy boundary and be able to identify what is within the galaxy disk and what is the stripped matter. In \cite{2022ApJ...925....4M}, the centroid of the brightest central region in the MUSE g-band on the map was used to designate the galaxy's center. Then, by masking the galaxy itself and, if any, its companions, the local background sky's surface brightness was determined. The stellar isophote that corresponds to surface brightness $1.5,\ 3,\ \text{and}\  5\sigma$ above the reported sky background level was then defined. Then, an ellipse was fitted to each isophotal level to define the galaxy boundary. The 5$\sigma$ contours are superimposed onto H$\alpha$ maps in Fig\ref{fig:A370_moretti} and \ref{fig:2744_moretti}.
\\

\begin{figure}
    \centering
    \includegraphics[width=\linewidth]{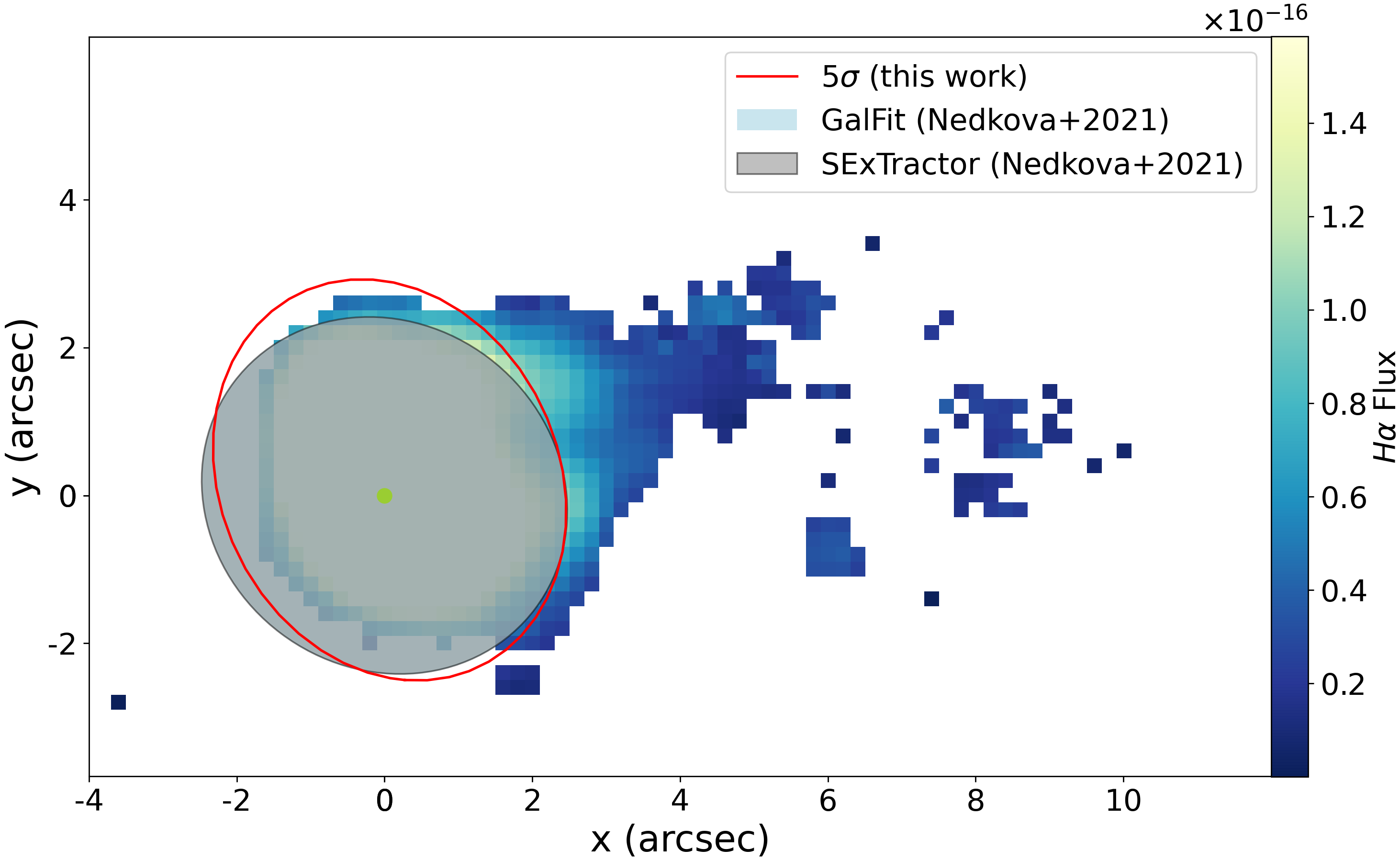}
    \caption{A370-01 H\(\alpha\) [$\mathrm{erg\ 
    cm^{-2}s^{-1}arcsec^{-2}}$] map of star-forming regions (See sec.\ref{subsec: spectral analysis}) over-layered by disk masks. The red ellipse denotes the disk boundary.
%    What we consider as the disk boundary in this work is denoted by the red ellipse. 
    SExTractor and GalFit (in this galaxy fully overlapping) disk position angles and ellipticity, 
%    (in this galaxy fully overlapping)
    described in Sec.\ref{disk_tail configuration}, are compatible with the 5$\sigma$ contours. The size of the comparison ellipse was adjusted to be the closest to our 5$\sigma$ contours for displaying purposes.}
    \label{fig:A370-01 Disk}
\end{figure}

We note that the definition of the parameters needed to distinguish between disk and tail might depend on the adopted band and algorithm. To assess the reliability of our measurements, we compared the ellipses used to define the disk masks to the ellipses obtained using the catalog of \cite{2021MNRAS.506..928N}, who measured structural parameters (e.g. position angle, inclination, and ellipticity) for Hubble Frontier Fields (HFF) fields (including A2744 and A370) that supplement the data release \((v3.9)\) of HFF-DeepSpace \citep{2018ApJS..235...14S}. We compared 
%Our comparison involved matching 
our results with their calculations of position angle and ellipticity (Fig.\ref{fig:A370-01 Disk}), which were obtained using two different algorithms, GALFIT and SExtractor, performed in the F$435$W band, which closely corresponds to the g-band. In the majority of cases, our findings are in good agreement with the results obtained by \cite{2021MNRAS.506..928N}. However, in three galaxies, we identified a discrepancy in the position angle that rendered it incompatible with the 5$\sigma$ contour. Upon visual inspection, we have reached the conclusion that the 5$\sigma$ contours align more closely with the information derived from the Hubble RGB images. Hence, in the following, we use our 5$\sigma$ contours to separate disks and tails in all galaxies.

\subsection{Spectral analysis}\label{subsec: spectral analysis}

The analysis of the MUSE data is extensively described in two key references, \cite{2017ApJ...844...48P} and \cite{2022ApJ...925....4M}. In a concise summary, MUSE spectra are corrected for the extinction caused by dust in our Galaxy.
%, as detailed in \cite{2017ApJ...844...48P}. 
The stellar-only component of the spectrum in every spaxel of each galaxy is derived using the spectrophotometric fitting code SINOPSIS (\citealt{2017ApJ...848..132F}) and is subtracted from each spaxel's spectrum to obtain the emission-only component. 
The stellar mass of each spaxel was obtained with SINOPSIS using the most recent update of the \cite{2003Bruzual} stellar population models \citep[see][for an outdated but still relevant description of the models]{2019Werle} with a \cite{2003PASP..115..763C} initial mass function. 

The gas kinematics, emission line fluxes, and associated errors are obtained using the software HIGHELF (Radovich et al. 2023, in preparation), which is based on the \href{https://lmfit-py.readthedocs.io/}{LMFIT} Python package, and fits Gaussian line profiles
using either one or two components. These fluxes are then corrected for intrinsic dust extinction using the \cite{1989ApJ...345..245C} extinction curve assuming an intrinsic Balmer decrement of \(H\alpha/H\beta  = 2.86\).

\begin{figure}
    \centering
    \includegraphics[width=\linewidth]{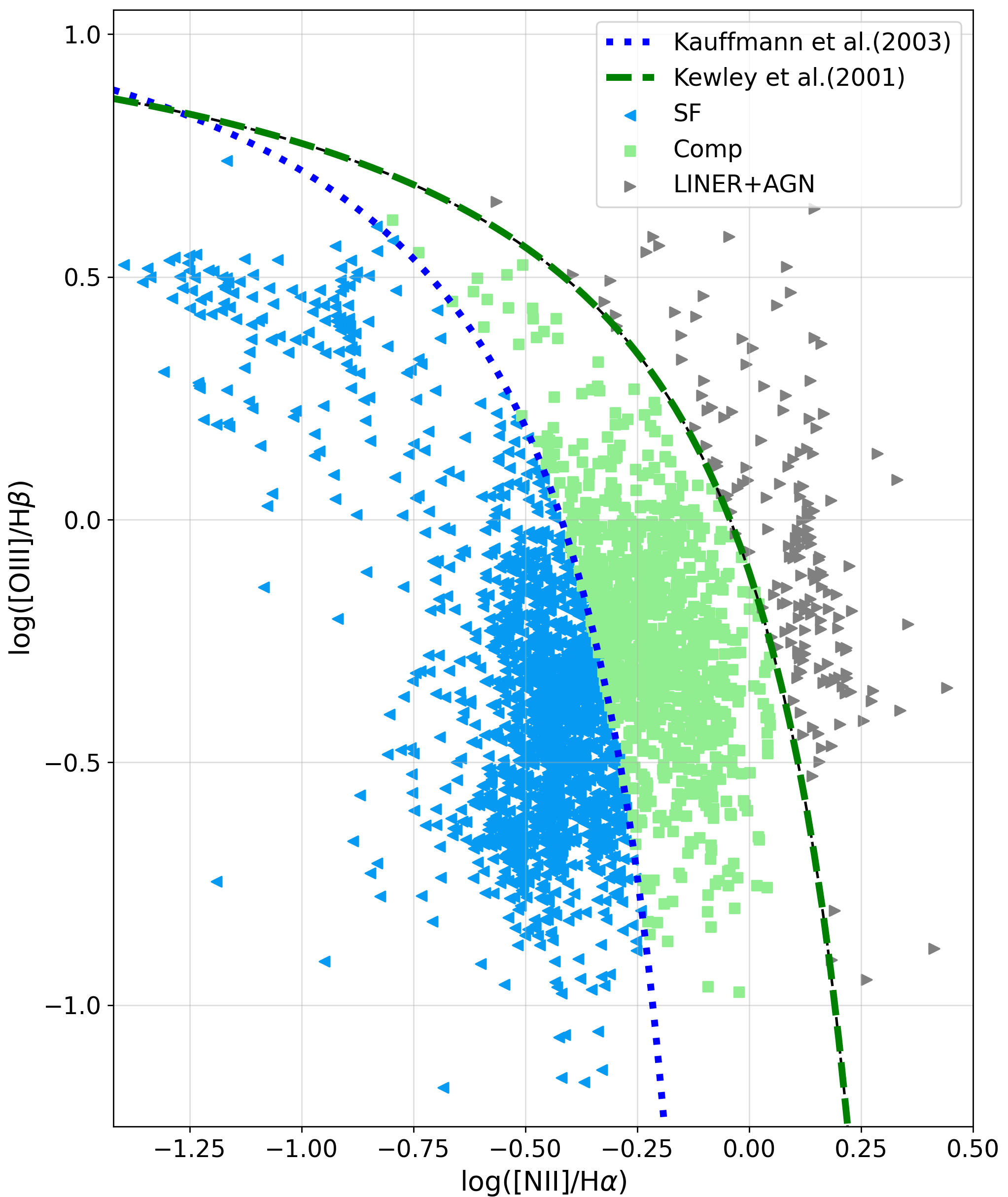}
    \caption{BPT diagram of all spaxels in the galaxy sample. They are classified as star-forming (SF), composite (Comp), and AGN/LINER. The Kauffmann and Kewley demarcation lines in [OIII]/H$\alpha$ vs [NII]/H$\beta$ BPT diagram are indicated by the dotted and dashed lines, respectively.  }
    \label{fig:BPT_ALL_SPX}
\end{figure}

 Diagnostic line ratios are employed to distinguish star-forming regions (which are assumed to be collections of unresolved \hii regions) from areas that are ionized by other mechanisms. Individual \hii regions have typical sizes of tens to hundreds of parsecs, hence are not resolved with our spatial resolution (i.e. $\sim 3.2$ kpc).

In the following, spaxels with signal-to-noise ratio S/N$\geq 2$ are characterized as star-forming using the \Nii/\Ha vs \Oiii/\Hb BPT diagram \citep{1981PASP...93....5B} according to the demarcation lines proposed by both \cite{2001ApJ...556..121K} and
\cite{2003MNRAS.346.1055K}, as seen in Fig.\ref{fig:BPT_ALL_SPX} where the spaxels of all galaxies are collectively plotted. Those spaxels that lie between these two lines of demarcation are labeled as composite while AGN and LINER regions are excluded from the rest of the analysis. For cases involving composite emissions, the observed line ratios might be influenced by additional ionization sources beyond star formation. These sources could potentially include AGN or LINER emissions \citep{2004Filho}, diffuse ionized gas \citep{2019Vale}, shocks \citep{2008Allen}, or elevated turbulence resulting from factors such as mixing of ISM and ICM due to the ram-pressure stripping \citep{2019MNRAS.482.4466P}. Consequently, spaxels classified as composite will be addressed individually in the subsequent spatially resolved analysis.

\begin{figure*}
    \centering
    \includegraphics[width=\linewidth,height=\textheight,keepaspectratio]{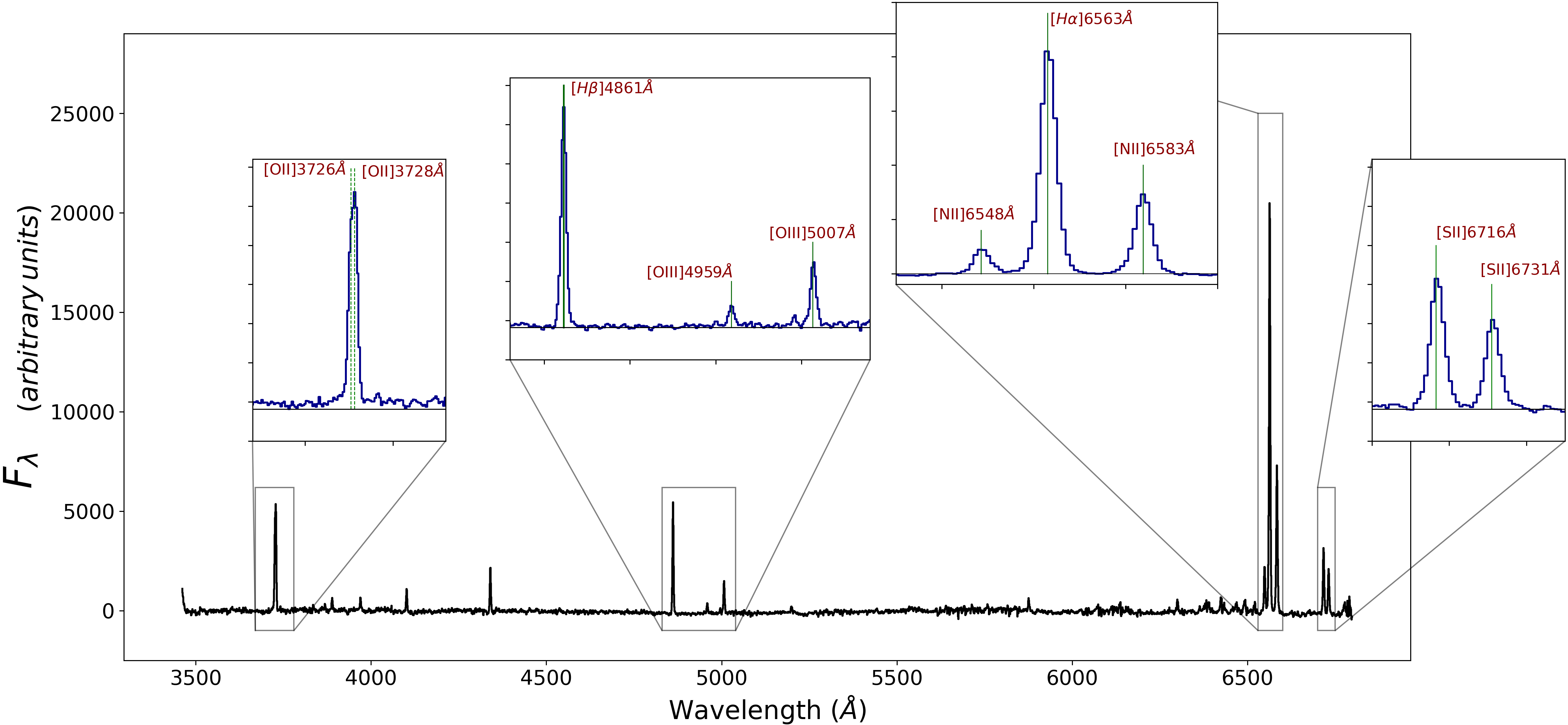}
    \caption{Rest-frame stacked spectrum for the disk of A370-01. Note that the stellar contribution is subtracted from each spaxel's spectrum, prior to stacking. Only the region of interest for this paper is shown, from 3500\AA to 6800\AA. In this range we have the essential nebular lines of [OII], [SII], [OIII], and H Balmer lines to measure the metallicity based on strong emission lines.}
    \label{fig:Stacked_spectra}
\end{figure*}

In addition to the spatially resolved analysis, in what follows we will also consider global quantities, always separating disks and tails. To accomplish this, we integrate the spectra of the disk and tail of each galaxy, separately.
Stacking spectra improves metallicity measurements of weak spectral features in low-mass or low-surface-brightness galaxies (e.g., A2744-03 and A2744-04). It also allows for statistical analysis of galaxy sample properties, particularly the global MZR. The results obtained by stacking are also more easily comparable to fiber or slit spectroscopic results (e.g. \citealt{2008A&A...488..463M}; \citealt{2010MNRAS.408.2115M}; \citealt{2013MNRAS.433.1425B}; and \citealt{2020MNRAS.491..944C})  and future works at different redshifts. As an example, 
A370-01 disk's stacked emission-only spectrum is shown in Fig.~\ref{fig:Stacked_spectra}. It contains the emission lines employed in the metallicity measurements.
 As an alternative method, we will also use the median metallicity in the disk and the tail separately. In both methods, we only consider spaxels powered by star formation and composite emission.

\section{Gas-Phase Metallicity Measurements}\label{sec: metallicity measurements}

In the realm of sub-solar metallicities, the most precise method for determining the gas phase oxygen abundance is often achieved by ascertaining the electron temperature of the nebula using the Te technique (also known as direct method), as described by \cite{1992MNRAS.255..325P} and \cite{2006A&A...448..955I}. This approach relies on the identification of faint auroral lines, which unfortunately remain largely undetectable within the average depth of observations conducted by the MUSE GTO and similar spectroscopic surveys. 

In this study, in the absence of direct temperature measurements, we effectively gauge the gas phase metallicity by using both a theoretical calibration through a photoionization model and an empirical calibration based on samples of \hii regions with direct method abundances. These strong line diagnostics ratios are used as indirect indicators. Though it is well known that different indicators provide different metallicities on an absolute scale, we choose to use two independent methods, based on different assumptions, in order to verify the solidity of our qualitative conclusions.

\subsection{Empirical Metallicity Measurement}

The first calibration we use was introduced by \cite{2004MNRAS.348L..59P}, P04 hereafter. This is an empirical method, calibrated via electron temperature measurements, that determines the oxygen abundance in \hii regions, particularly with an eye to their application to the analysis of star-forming galaxies at high redshift. It is based on the O3N2 index, defined as O3N2 $\equiv$ $\log$[ (\Oiii/H$\beta$)/([\Nii/H$\alpha$ )] according to the formula:
\begin{equation}\label{p04_Equation}
    12 + \log (O/H) = 8.73 - 0.32 \times \text{O3N2}
\end{equation}

Equation \ref{p04_Equation} is valid only for O3N2 $< 1.9$, corresponding to a minimum metallicity of $\sim 8.1$. Spaxels with O3N2 $\geq 1.9$ must be excluded in our analysis, and represent $< 5$\% of our set of spaxels in each galaxy, except in A2744-04 in which they are $\sim 30$\% of usable spaxels. Notably, a probable systematic error in the super-solar metallicity regime affects the calibration \citep{2002RMxAC..12...62S}, which is empirical in nature and anchored to the abundances of 137 \hii regions, six of which are determined through photoionization models and the rest by the direct method. Additionally, it should be emphasized that since [N \textsc{ii}] and [O \textsc{iii}] come from ions with a significant difference in ionization potential, changes in the ionization parameter, which are uncalibrated, consistently influence this diagnostic ratio (see, for instance, \citealt{2015ApJ...798...99B}).

However, the O3N2 index offers the advantage of minimizing uncertainties resulting from flaws in flux calibration or extinction correction due to its reliance on line ratios that are close in wavelength and exhibit a monotonic change with metallicity. Consequently, it serves as a valuable calibration for measuring the metallicities of both the A370 and A2744 clusters. This calibration can also be applied to slightly more distant clusters observed with MUSE. Moreover, it is widely used at $z>0$, hence it is convenient to compare with future works.\\ 

Using Monte Carlo simulations, the inferred P04 metallicity uncertainties are calculated by randomly perturbing (under the assumption of a Gaussian noise distribution) all recorded line fluxes by their measurement errors 1000 times. Thus, the metallicities are the median of these samples and the reported errors are their standard deviations. Nevertheless, we point out that since the systemic contributions are not taken into account, these values likely underestimate the actual uncertainty.

\begin{figure*}

    \centering
    \includegraphics[width=\linewidth]{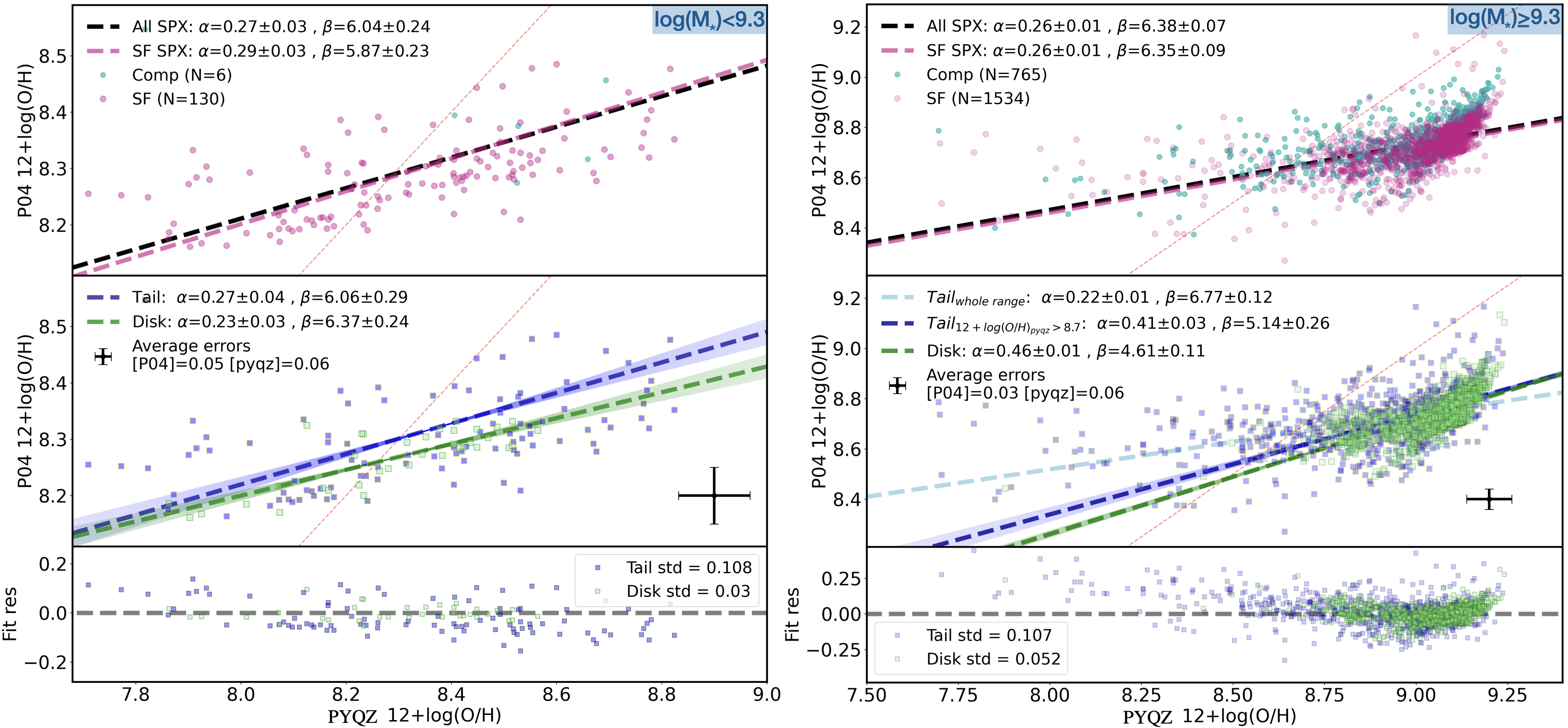}
    \caption{Comparison of the metallicity measurements of all spaxels in all galaxies of our sample for the two different metallicity diagnostics. In both right (high-mass galaxies, $\log(M_\star/M_{\odot}) \geq 9.3 $) and left (low-mass galaxies, $\log(M_\star/M_{\odot}) < 9.3 $) plots, the behavior of star-forming (SF) and composite (Comp) spaxels can be seen in the top panels. Their relationship is modeled via a linear approach and least square fit where \( P04_{12 + \log (O/H)} = \alpha \times \text{\textsc{pyqz}}_{12 + \log (O/H)} + \beta\). The dotted red line represents the one-to-one relation while the black and the pink dashed lines are SF+Comp (all spaxels) and SF linear fits, respectively. Middle panels of both plots: disk (green) and tail (blue) spaxels are linearly modeled, distinctly. The error bars are the metallicity uncertainties described in Sec.\ref{sec: metallicity measurements}. The bottom panel of both plots: residuals from the overall fit (black line in top panel) with standard deviation (std) values in the label.}
    \label{fig:pyqz_p04 Comparison High-Mass}
\end{figure*}

\subsection{\textsc{pyqz}}

For a specific set of emission line fluxes, the \textsc{pyqz} python library (\citealt{2013ApJS..208...10D}; \citealt{2015MNRAS.450.2593V}) concurrently delivers the gas metallicity and ionization parameter ($\log \, q$) values. In order to interpolate the $\log \, q$ and 12+log(O/H) values, a finite number of diagnostic line ratio grids was produced using the MAPPINGS code \citep{1993ApJS...88..253S, 2013ApJS..208...10D}. \textsc{pyqz} creates N (i.e. N=400 set as default) randomly chosen sets of emission-line fluxes, Gaussian distributed in accordance with the given flux errors, considering the available emission lines of each spectrum. The algorithm calculates the metallicity and the ionization parameter for each set, then performs a two-dimensional Gaussian kernel density estimate to create the probability density function (PDF) from the discrete distribution of the metallicity and ionization parameter values. %Furthermore, 
To determine the metallicity uncertainty \citep{2015MNRAS.450.2593V}, the algorithm propagates the PDF according to the flux measurement uncertainty and defines the metallicity errors as the \(1\sigma\) contours of the PDF.\\
% \textsc{pyqz}'s grids are flat and unwrapped, thus they allow us to disentangle the effects of log(Q) and 12+log(O/H) on the emission line ratios.\\

Here, we utilize the same modified version of \textsc{pyqz} that was used by \cite{2020ApJ...895..106F}. This version makes use of MAPPINGS IV to produce diagnostic line ratio grids and covers the range \(7.39 \leq 12 + \log(O/H) \leq 9.39\), which works well even in the extremely high metallicity range, and the \(6.5 \leq \log q \leq 8.5 \) ionization parameter range. This \textsc{pyqz} version covers a wider range of line ratios with respect to the latest official version (i.e. v0.8.4), thus allowing us to recover metallicities for a larger number of spaxels.

We employ the [O \textsc{iii}]4959,5007\r{A}, [N \textsc{ii}]6548,6583\AA, and [S \textsc{ii}]6716,6730\AA doublets as inputs since they offer a solid distinction between log(q) and 12+log(O/H) and minimize the influence of the reddening correction uncertainty.  
%process of applying reddening corrections is straightforward. 
We utilize the strongest line flux of each of the doublets
[O \textsc{iii}]4959,5007\AA and [N \textsc{ii}]6548,6583\AA
%in our measurements 
to reduce measurement uncertainty and calculate the amplitude of the weaker line using the ratio of the relative Einstein coefficient. As a result, the necessary emission line ratios are ($1.3\times$\Oiii)/([S \textsc{ii}]$\lambda6716$+[S \textsc{ii}]$\lambda6730$) and ($1.3\times$\Nii)/([S \textsc{ii}]$\lambda6716$+[S \textsc{ii}]$\lambda6730$).\\

Furthermore, the \textsc{pyqz} algorithm reports the uncertainty based on flux errors. It also suffers from not taking into account the systematic uncertainty, but \cite{2020ApJ...895..106F} offered an estimation of systematic error in relation to the \textsc{pyqz} model grid uncertainty. In a nutshell, a model uncertainty of 0.1 dex results in a systematic error of 0.05 dex on the metallicity estimate for the highest metallicities (12 + log (O/H) = 9.39), while for the lowest metallicities (12 + log (O/H) = 7.39), the error can extend up to 0.3 dex. In our sample, the majority of oxygen abundances exceed 8.95. Consequently, considering the aforementioned systematic uncertainty associated with these abundances, we incorporate an additional 0.05 dex to the measured metallicity errors.

% \subsection{Error Estimation}\label{error_estimation}

% \\ 

\begin{figure*}
    \centering
    \includegraphics[width=\linewidth]{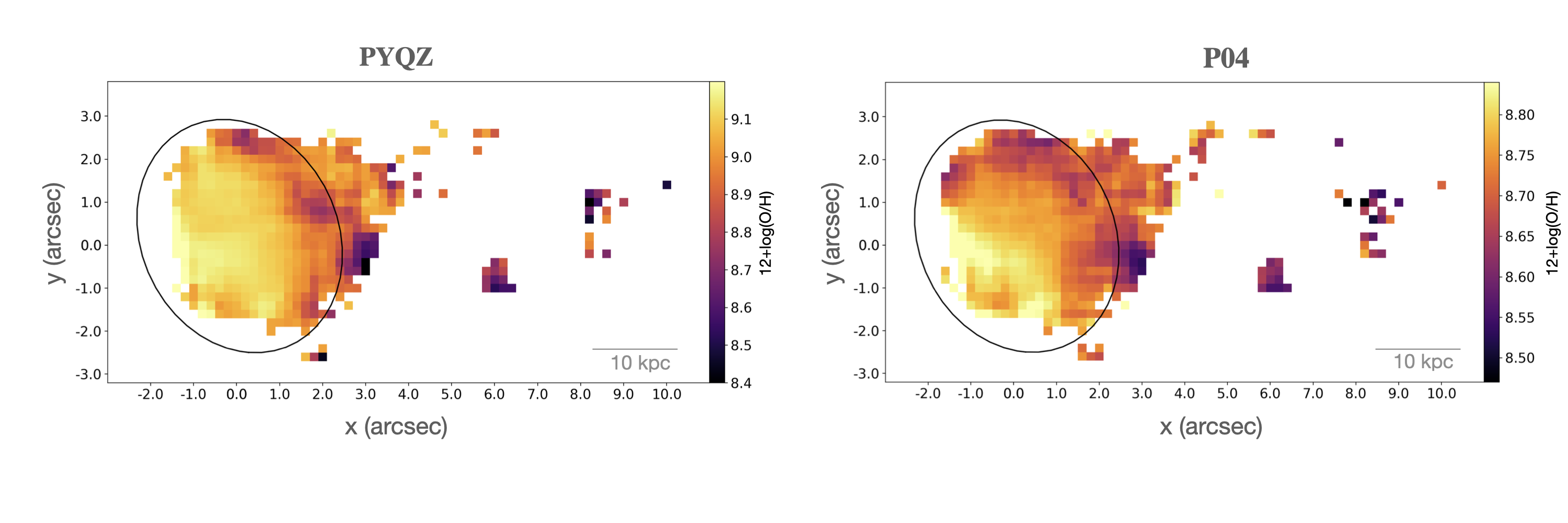}
    \caption{A370-01 metallicity maps. The black ellipses identify the \(5\sigma\) contour over the background which we recognize as the disk boundary. The left panel shows the 12+log(O/H) map calculated by \textsc{pyqz} and the right corresponds to P04. Since the two methods employ distinct emission lines to compute metallicity (as mentioned in Sec.\ref{method_comparison}), they do not necessarily share the same usable spaxels. 
    Additionally, as explained in Section \ref{sec: metallicity measurements}, intrinsic variations in the estimated absolute values using the two methods render it impractical to present maps with identical metallicity ranges as a colorbar.}
    \label{fig:A370_01 Met Map}
    
\end{figure*}
% \\

\subsection{Method Comparisons}\label{method_comparison}

Even when using the same diagnostic ratios, different calibrations can provide absolute metallicity estimates that vary by as much as 0.6 dex \citep{2012ApJ...756L..14P, 2012MNRAS.426.2630L}.  Regardless of the absolute abundance scale problem (e.g. \citealt{2008ApJ...681.1183K}), which is beyond the scope of this work to resolve, Fig.\ref{fig:pyqz_p04 Comparison High-Mass} shows a strong correlation ($\emph{p}>.6$) between the two metallicity calibrations, for the spaxels with both metallicity estimates and both for galaxies with mass below and above $M_{\star}=10^{9.3} M_{\odot}$. Despite these strong correlations, minor discrepancies may arise in their behavior for specific scenarios (e.g., see sections \ref{sec: met gradient} and \ref{subsec: MZR}).\\

We note that, as shown in the right plot's upper panel of Fig.\ref{fig:pyqz_p04 Comparison High-Mass}, the correlation between the two metallicity indicators does not change when including or excluding composite spaxels. Moreover, the estimates given by the two indicators differ the most in the tails (middle and bottom panels in Fig.\ref{fig:pyqz_p04 Comparison High-Mass}): in fact, the tail points are the most deviant from the linear fit with $\sigma \sim 0.1$ dex. More importantly, the tail and disk fit relations for high-mass galaxies have different slopes considering the whole range of metallicity (light-blue dashed line), with negligible uncertainties as shown in Fig.\ref{fig:pyqz_p04 Comparison High-Mass}. This discrepancy in the linear fit slopes is primarily attributed to the occurrence of low \textsc{pyqz} metallicity estimations in certain tail spaxels, where the value of $12+\log(O/H)_{\text{\textsc{pyqz}}}$ is less than 8.7. However, if we only consider tail spaxels with metallicities falling within the range covered by disk metallicities (i.e. 12+$\log$(O/H)$_{\text{\textsc{pyqz}}}>$ 8.7), the slopes of the fits for disk and tail become more similar.
 
For this reason, and taking into account the additional factors mentioned earlier, in the following we will exploit both indicators, to assess whether results depend on the chosen indicator.

\section{RESULTS}\label{sec: results}
%GAS PHASE METALLICITY MEASUREMENTS} 

\begin{figure*}[h!]
    \centering
    \textbf{P04 Metallicity Maps}
    \includegraphics[width=\textwidth,height=0.94\textheight,keepaspectratio]{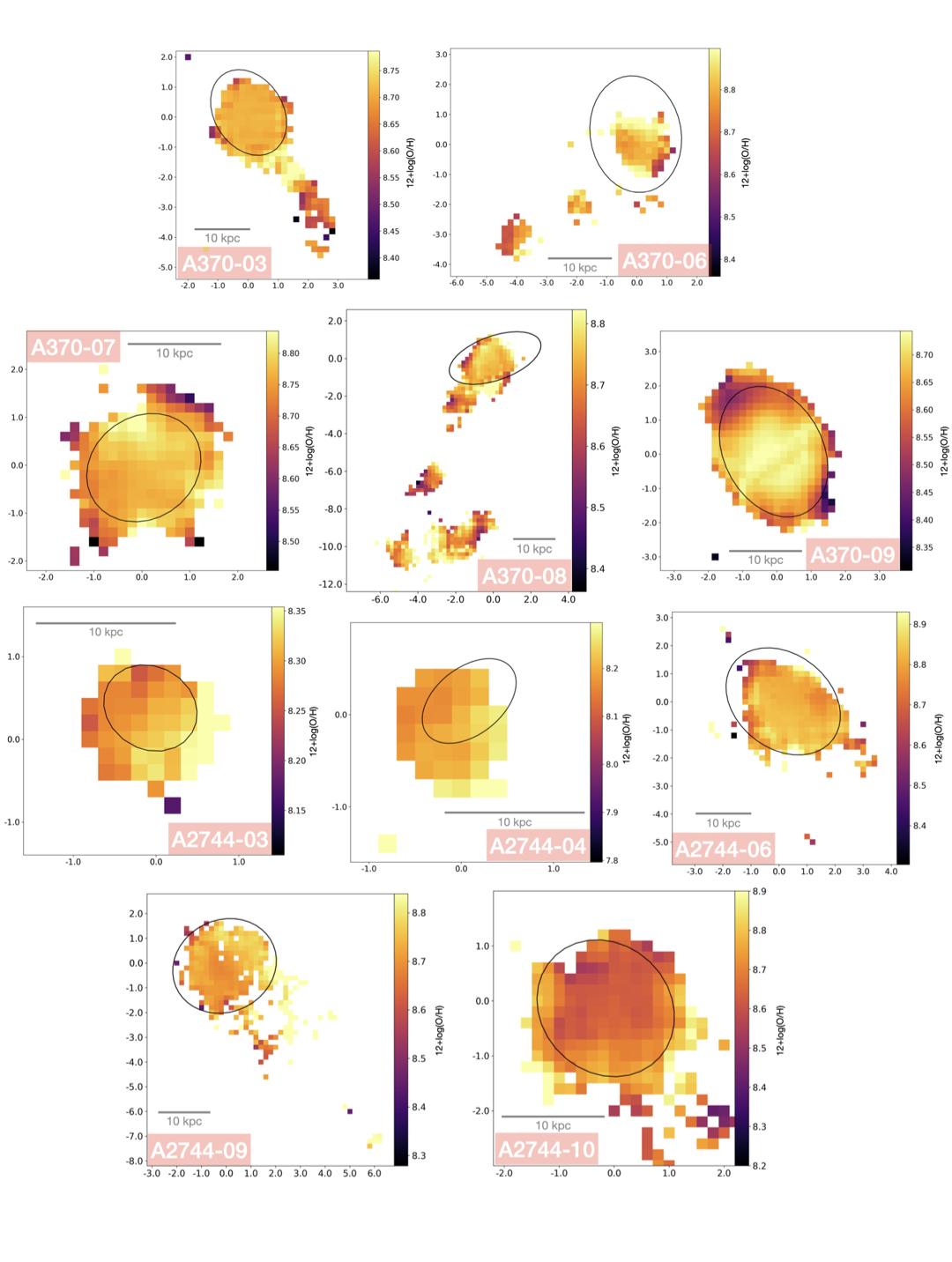}
    \caption{P04 metallicity maps of 10 RPS galaxies in A370 and A2744 clusters. Each map’s black outlined ellipse represents the 5$\sigma$ contour, which is the boundary of the galaxy disk (see Sec.\ref{disk_tail configuration}). Each plot also shows a scale corresponding to 10kpc at the cluster redshift.
}
    \label{fig:p04_met_map}
\end{figure*}

We determine the metallicity value for each spaxel that meets the aforementioned BPT and signal-to-noise ratio (SNR$>$2) criteria. In Fig.\ref{fig:A370_01 Met Map}, we compare the metallicity map obtained by \textsc{pyqz} (left panel) and P04 (right panel) of the galaxy A370-01, while Fig.\ref{fig:p04_met_map} and Fig.\ref{fig:pyqz_met_map} cover rest of the sample. As these figures demonstrate, we can measure the gas metallicities also in regions outside of the stellar disk, in some cases only close to the disk outskirts and in other cases far from the disk, up to $\sim 50$ kpc away (e.g. A370-08).

\pagebreak

\begin{figure*}[h!]
    \centering
    \textsc{pyqz} \textbf{Metallicity Maps}
    \includegraphics[width=\textwidth,height=1.\textheight,keepaspectratio]{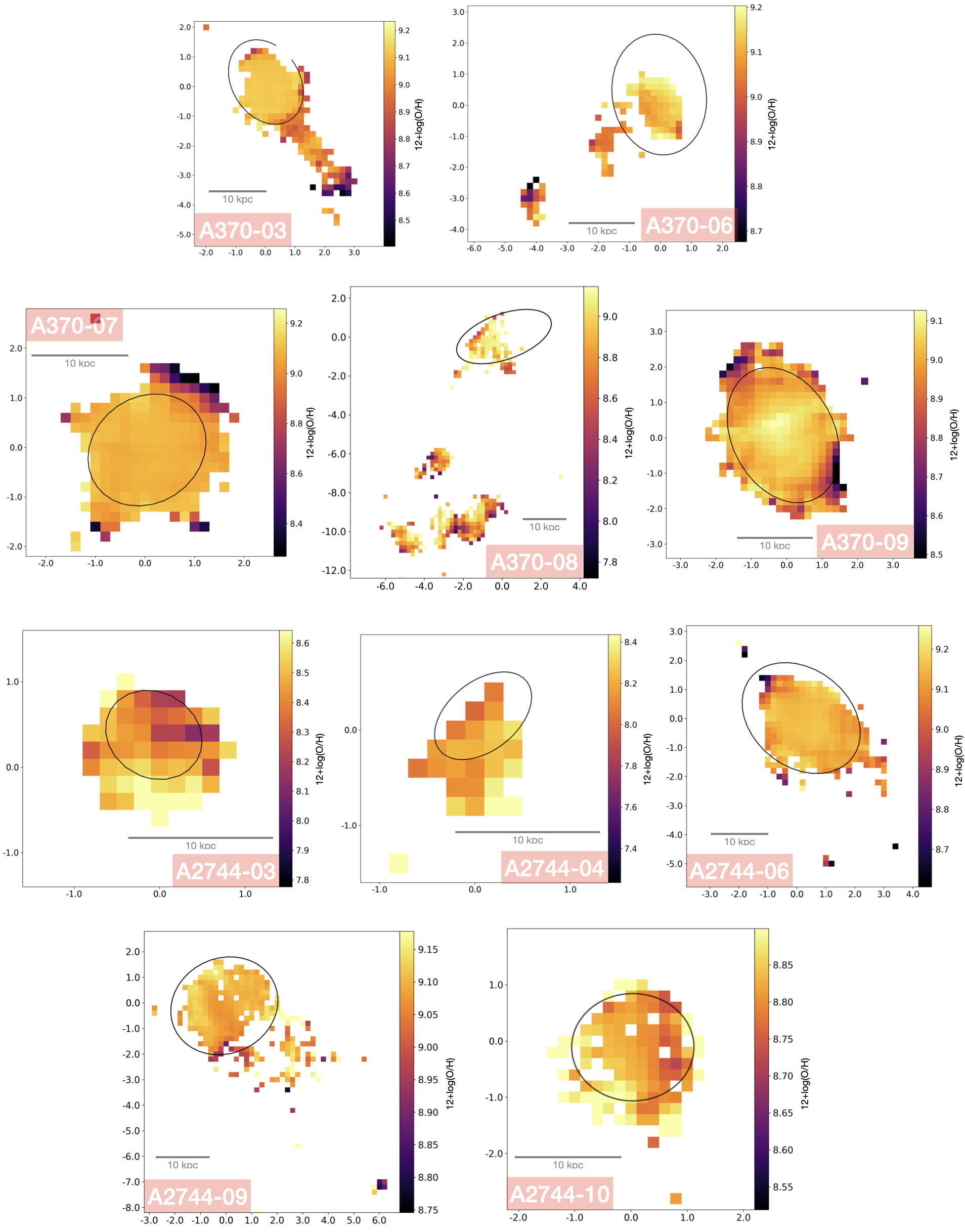}
    \caption{ \textsc{pyqz} metallicity maps of 10 RPS galaxies in A370 and A2744 clusters. The 5$\sigma$ contour, which is the boundary of the galaxy disk (see Sec.\ref{disk_tail configuration}), is depicted as a black ellipse on each map. Additionally, a scale for 10kpc at the cluster redshift is included for each plot.}
    \label{fig:pyqz_met_map}
\end{figure*}

\subsection{Metallicity trend with projected distance}\label{sec: met gradient}

In order to investigate the behavior of oxygen abundances as we move further from the central region of galaxies to their tails, we compute the metallicity as a function of angular distance. It is important to note that while a deprojection procedure for the disk could be accomplished using the disk axial ratio, the deprojection of the tail is impossible due to the unknown orientation. Our study primarily centers on comparing metallicities in both the disk and tail regions, and as such, we exclusively employ angular and projected distances throughout the entirety of the paper.

We present the metallicity gradient (projected) of all galaxies in our sample in Fig. \ref{fig:Metgrad}, where we differentiate between star-forming and composite spaxels. The error bars represent the $1\sigma$ percentiles, while the median values reflect the median metallicity of each radial bin, considering both star-forming and composite regions. Below, we present a list of several observed features that can be noted upon initial examination.

%\begin{figure*}
%    \centering
    %\includegraphics[width=\linewidth]{A370_01_Met_Gradient.png}
 %   \caption{A370-01 metallicity gradient computed with \textsc{pyqz} (left panel) and P04 (right panel).
    %(with $\log(M_\star/M_\odot)=10.6$) is computed using two different methods: pyqz (left panel) and P04 (right panel). 
 %   Disk and tail spaxels are color‐coded with green and blue, respectively. The median uncertainties for disk and tail spaxels are provided in the right corner of the panels. Dark triangles represent composite spaxels (Comp) and light squares are star-forming (SF). The bold plus markers are median 12+log(O/H) values in each angular distance bin while associated error bars (and shades) represent 1$\sigma$ percentiles.}
%    \label{fig:A370_01 met gradient}
%\end{figure*}

\begin{enumerate}[label=\roman*.]

\item The P04 and \textsc{pyqz} metallicity trends qualitatively agree with one another, except in A370-03 where they differ only for the part of the tail that is closer to the disk, and in A2744-09 where the tail's trend is flat with respect to the disk using \textsc{pyqz}, while the metallicity increases in the tail for P04. 
%As highlighted in Section \ref{method_comparison}, \textsc{pyqz} and P04 employ distinct emission line ratios for the determination of metallicity, and thus may not share the same set of usable spaxels.

\item  No large systematic difference in the metallicity gradient behaviors of SF and composite spaxels is seen since they mostly follow the same path, except for the fact that in a few of the disks, the most deviant points are composite (typically towards low metallicities). A separate fit for composite and star-forming spaxels is only possible for the tails of two galaxies, A370-03 and A370-09, where the best fit of star-forming and composite spaxels is consistent within the uncertainties.
The composite deviant points in the disks (Fig.\ref{fig:BPT_MAPS}) are found either in the region of the disk where the tail begins (thus they might actually be already a part of the tail seen in projection onto the disk) or at the shock front caused by ram pressure, right at the leading edge of the disk. This is evident in cases like A370-01, A370-03, A2744-06, A2744-09, and A2744-10. A370-06, the only one with an AGN in the sample, and A370-08 stand out instead for having the whole disk dominated by composite spaxels.

\item Metallicity trends in the disks are typically negative or flat with two positive 
exceptions, A370-06, and A2744-10. Note that A370-06, the one with the AGN, has a large number of AGN-powered spaxels along an extended ionization cone \citep[][and Fig.\ref{fig:BPT_MAPS}]{2022ApJ...925....4M}, which are likely to affect our capability to assess the profile.

\item  In most cases, the tail metallicity declines with distance along the tail and
is generally lower than in the corresponding disk, with three notable exceptions: A2744-04, A2744-09, and A2744-10. Moreover, the slope in the tail is often not the same as the slope in the disk (e.g. A370-03). This may be the result of projection effects since we are unable to establish the actual separation of tail spaxels from the galactic center.

\item  Overall, the global metallicity gradient (disk + tail) is negative or almost flat (except for A2744-10 and P04 estimate of A2744-09 which have sharp positive slopes as already discussed above). 

\end{enumerate}

\setcounter{figure}{9}
\begin{figure*}[h!]
    \centering
    % \figurenum{10}
    % \textbf{}
    \includegraphics[width=\textwidth,height=0.9\textheight,keepaspectratio]{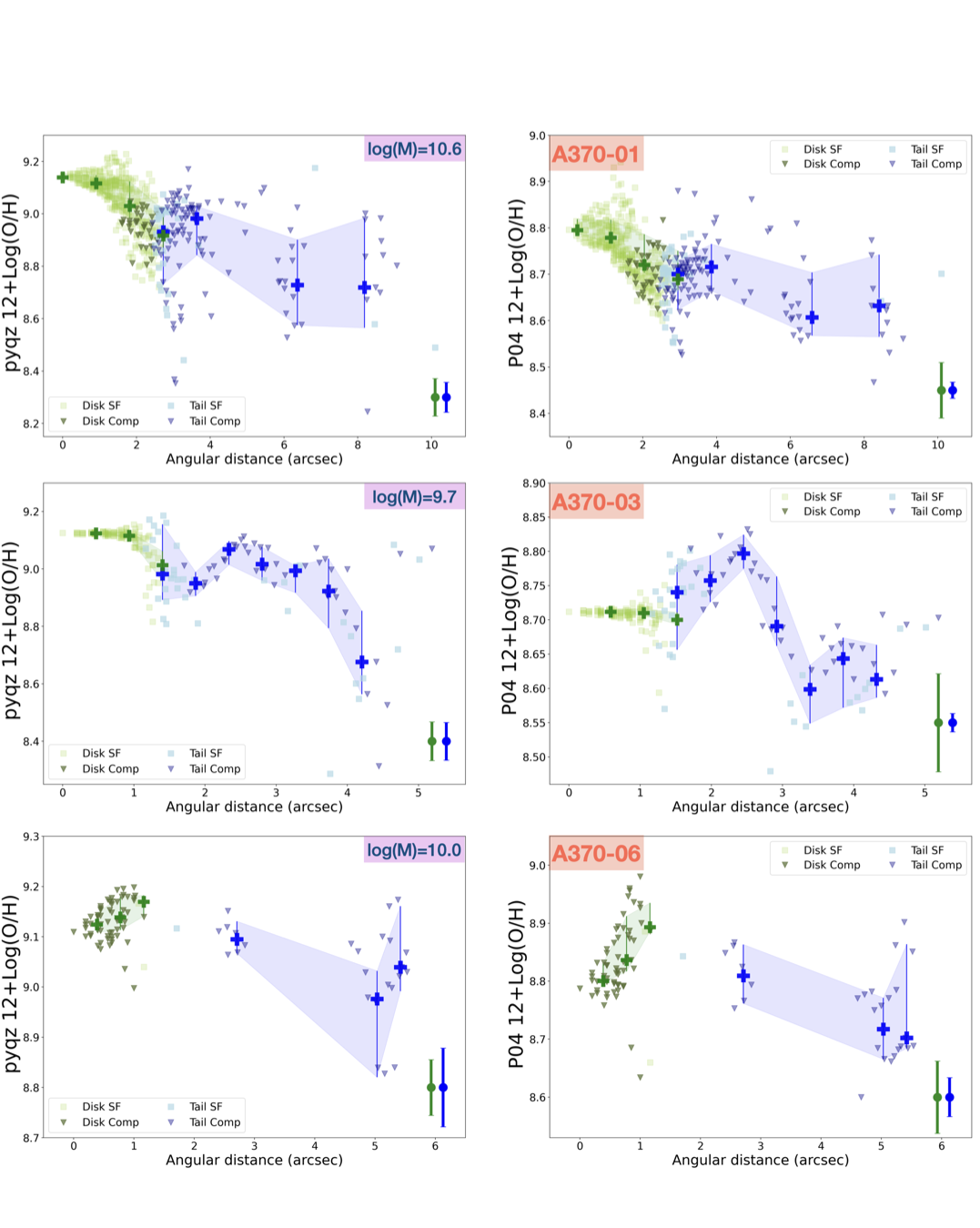}
    \caption{(continued)}
    \label{fig:Metgrad1}
\end{figure*}
\setcounter{figure}{9}
\begin{figure*}[h!]
    
    \centering
    % \textbf{}
    \includegraphics[width=\textwidth,height=0.95\textheight,keepaspectratio]{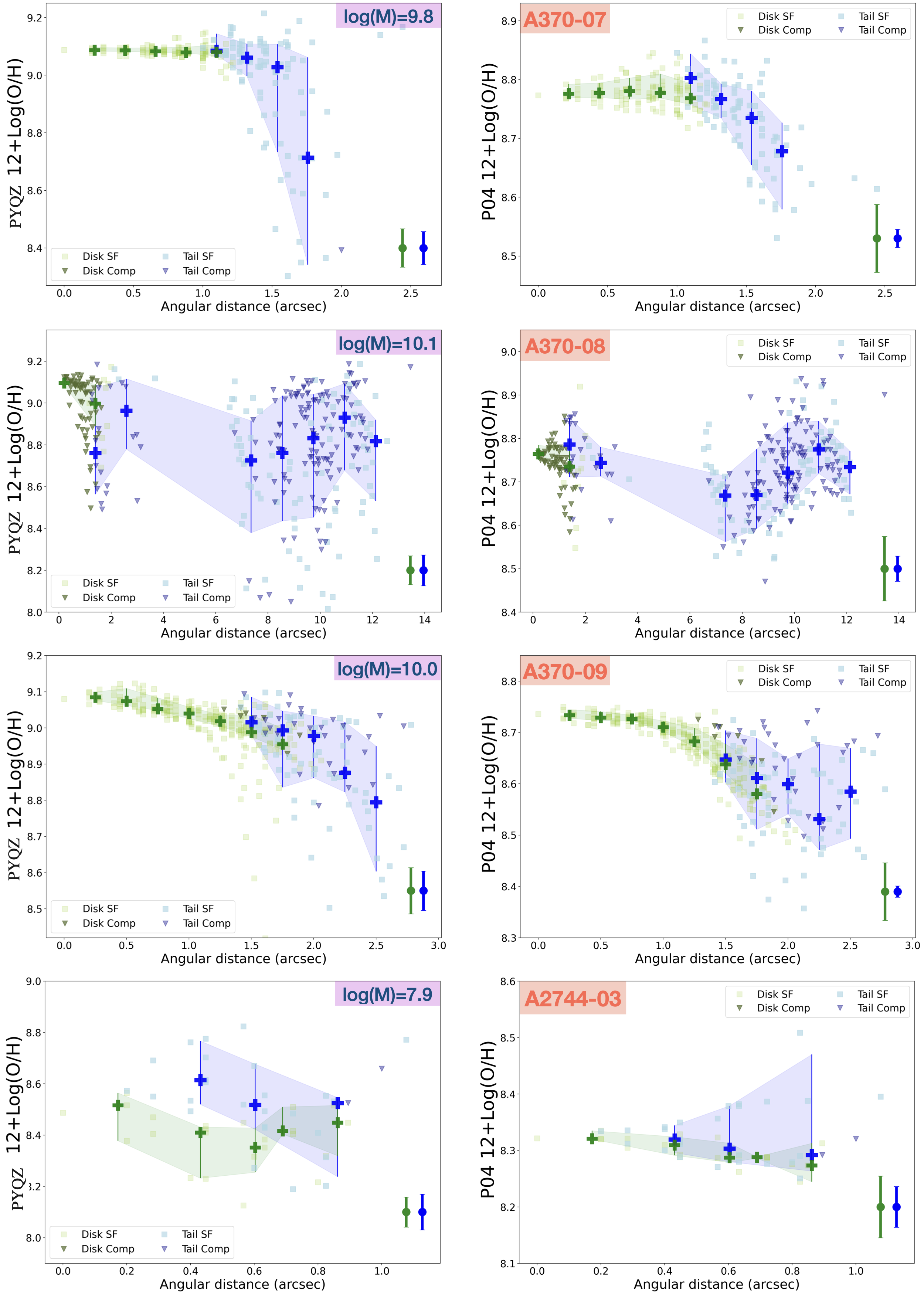}
    \caption{(continued)}
    \label{fig:Metgrad2}
\end{figure*}

\setcounter{figure}{9}
\begin{figure*}[h!]
    % \figurenum{10}
    \centering
    % \textbf{}
    \includegraphics[width=\textwidth,height=0.93\textheight,keepaspectratio]{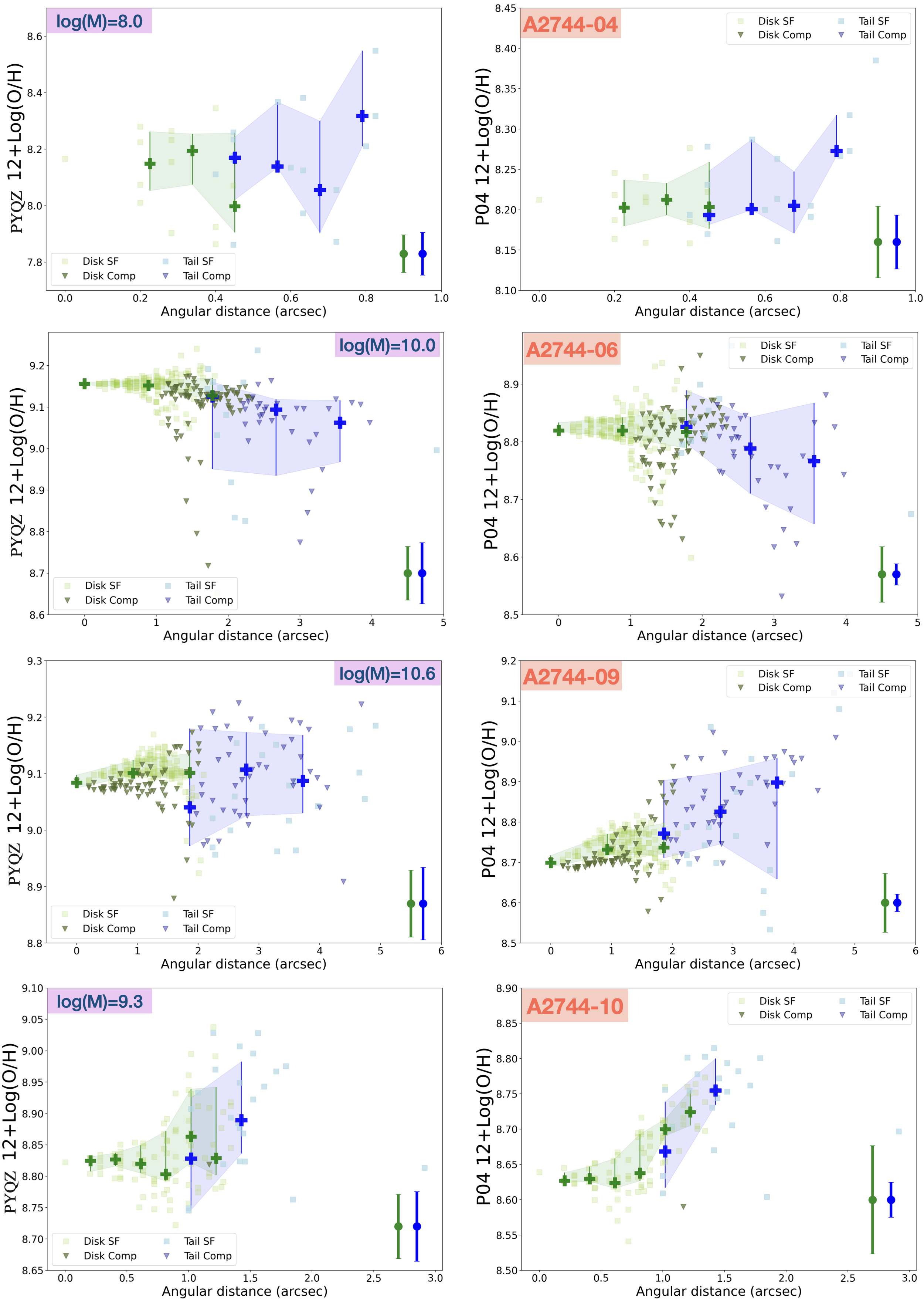}
    \caption{metallicity gradients were computed using two different methods: \textsc{pyqz} (left panel) and P04 (right panel). Disk and tail spaxels are color‐coded with light and dark green, and light and dark blue, respectively. Triangles represent composite spaxels (Comp) and squares are star-forming (SF). The bold plus markers are median 12+log(O/H) values in each angular distance bin while associated error bars (and shades) represent 1$\sigma$ percentile. Also, the stellar mass of galaxies (M)is written in the left panels in solar mass units. }
    \label{fig:Metgrad}
\end{figure*}

\begin{figure*}[h!]
    \centering
    % \textbf{}
    \includegraphics[width=0.96\linewidth]{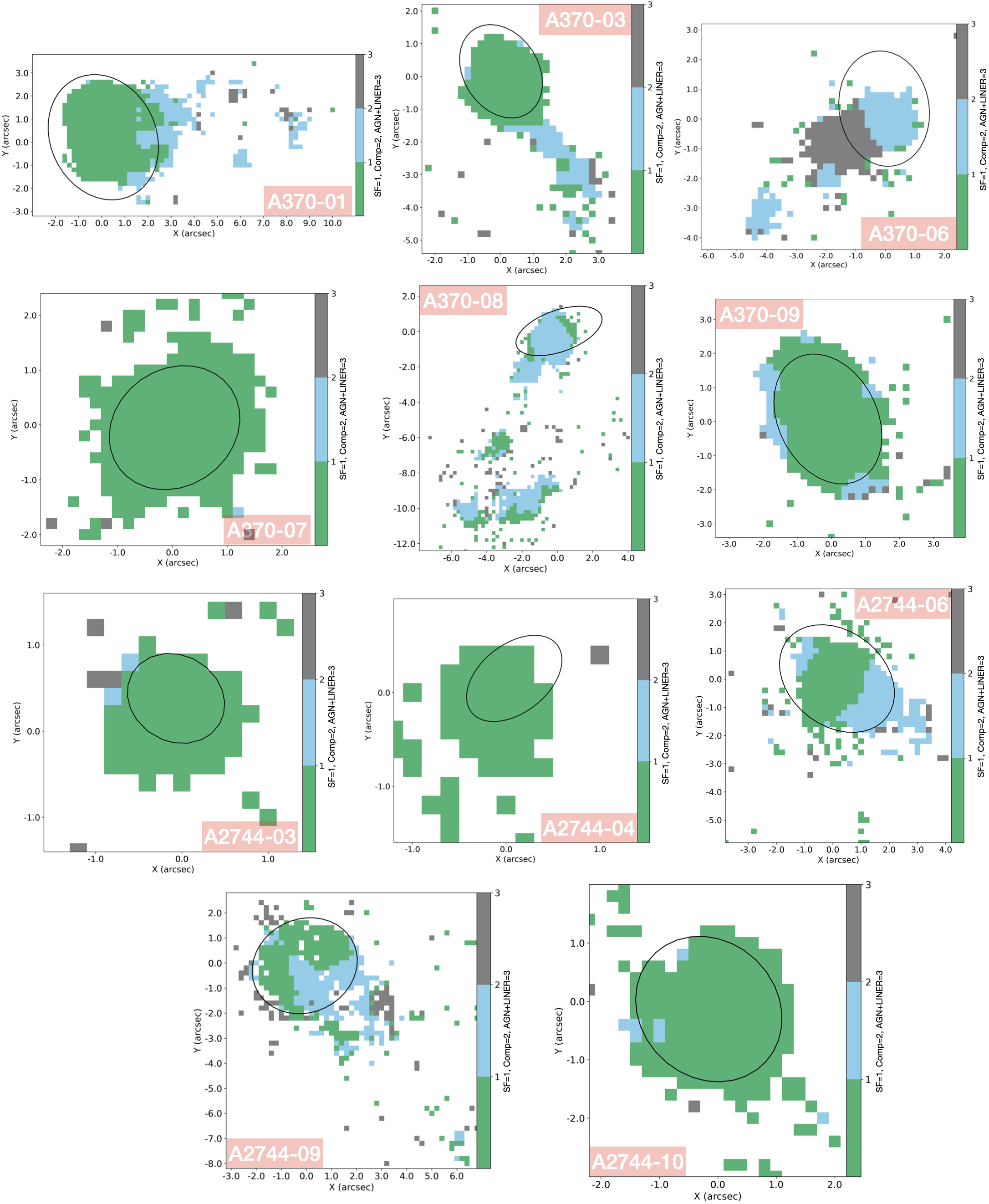}
    \caption{BPT Maps of 11 RPS galaxies in A370 and A2744 clusters. Green spaxels are those dominated by star formation, blue ones are classified as composite, and gray spaxels lie either in the LINER/shocked or AGN regions.}
    \label{fig:BPT_MAPS}
\end{figure*}

In addition, ionization characteristics (composite vs. SF) of tails and disks are not necessarily coupled (see fig.\ref{fig:BPT_MAPS}). In 5 galaxies, these properties agree, being both tail and disk dominated ($>95\%$) either by SF (A370-07, A2744-03, AA2744-04, A2744-10) or by composite emission (A370-06). However, in 3 other galaxies, the emission in the tail has probably a different origin from that in the disk 
% MG: this is the original version 
%\citep[see low-redshift studies],[]{2019MNRAS.482.4466P,2021ApJ...922..131T}.
% below how i modified
\citep[see low-redshift studies][]{2019MNRAS.482.4466P,2021ApJ...922..131T}.
For instance, in A370-01 the tail is mostly composite while the disk is mostly SF. Also, there are cases such as A370-08 where the disk is dominated by composite emission, while in the tail numerous spaxels are found both in the composite and in the SF regime.  

It is interesting to assess whether the observed trends depend on galaxy stellar mass. The two lowest mass galaxies, A2744-03 and A2744-04, (of the order of $\log(M_\star/M_{\odot}) \sim 8$) have trends consistent with a flat metallicity gradient across disk and tail, although with a large uncertainty and a few points. Also, considering the proximity of the disk angular size of these galaxies to the observation's resolution limit, this might be simply a resolution effect. The only galaxy in which the tail metallicity appears to be consistently higher than in the disk (\textsc{pyqz}), or at least to have a growing outward profile (P04), is A2744-10, with a mass $\log(M_\star/M_{\odot})=9.3$. As mentioned above, \textsc{pyqz} and P04 results do not agree for A2744-09 and it is, therefore, hard to draw any conclusion for this galaxy. For all the other galaxies (all with $\log(M_\star/M_{\odot}) \geq 9.7$) the tail reaches metallicities lower than anywhere in their disk.

Our findings are in line with the two existing studies to date at redshifts $\sim 0$, as reported in \cite{2019MNRAS.485.1157B} and \cite{2021ApJ...922L...6F}. They examined the projected tail star-forming clumps of, respectively, 1 and 3 RPS galaxies with $\log(M_\star/M_{\odot})$ equal to 10.55, 10.96, 11.21, and 11.50. These studies revealed a maximum metallicity difference of $\sim0.5$ dex between the nearest and farthest parts of the tails from the disk by using the same modified version of \textsc{pyqz} used in this paper. It should be noted that the projected tail lengths observed in the sample of \cite{2019MNRAS.485.1157B} and \cite{2021ApJ...922L...6F} are approximately 20-50 kpc.

In contrast, our own typical projected tail lengths are equal to or less than 20 kpc (with only one exception of $\sim 50$ kpc), and at considerably higher redshifts. Also, due to the differences in the approach employed, with our analysis focusing on the metallicity of individual spaxels\footnote{The MUSE spatial resolution is not sufficient to identify individual star-forming clumps in z$\sim$0.3 RPS galaxies.} while their work examined integrated spectra of star-forming clumps, a direct comparison with \cite{2021ApJ...922L...6F} is not feasible. \\

\subsection{Mass-Metallicity Relation}

 In this section, we first discuss the spatially-resolved mass-metallicity relation of each galaxy, distinguishing disk and tail spaxels. We then analyze the global mass-metallicity relation separately for disks and tails, to investigate whether the tail metallicity is correlated with galaxy mass as the disk metallicity is.
\subsubsection{Spatially-Resolved Mass-Metallicity Relation}

 Fig.\ref{fig:MZR_TOT} presents the resolved mass-metallicity relation of the whole sample with few exceptions. Note that SINOPSIS (see Section \ref{subsec: spectral analysis}) cannot calculate the surface density of stellar mass for galaxies A2744-03 and A2744-04, since their spectra do not contain any information about the stellar continuum. Moreover, the surface density of stellar mass for A2744-10 is not yet available. This figure illustrates that in both metallicity diagnostics, P04 and \textsc{pyqz}, the metallicity rises as we approach spaxels with higher stellar mass surface density, which is consistent with other findings on rMZR \citep[e.g.,][] {2012ApJ...756L..31R,2022A&A...661A.112Y}. The tail of the galaxy does not reach the highest surface mass density reached in the disk and at any given mass density its median metallicity trend follows that of the disk, with values that are similar or slightly lower. 

 The rMZR of the disk is mostly flat or positive (only A370-01 and A370-09) and one case slightly negative (A2744-06). The stellar-mass surface densities of tail spaxels consistently populate the lower end of the density range observed in the disks, and they do not necessarily illustrate the same trend as the disks regardless of their parent galaxy stellar mass values (see Tab.\ref{tab:P04 stack+med table+z+M}). Moreover, the presence of spaxels in the tails with significantly lower metallicities than any disk spaxel of similar surface mass density is a noteworthy finding. This may indicate different enrichment processes in the tails compared to the disk, as previously mentioned.\\

\begin{figure*}[h!]
    % \figurenum{12}
    \centering
    \includegraphics[width=\linewidth,height=0.94\textheight,keepaspectratio]{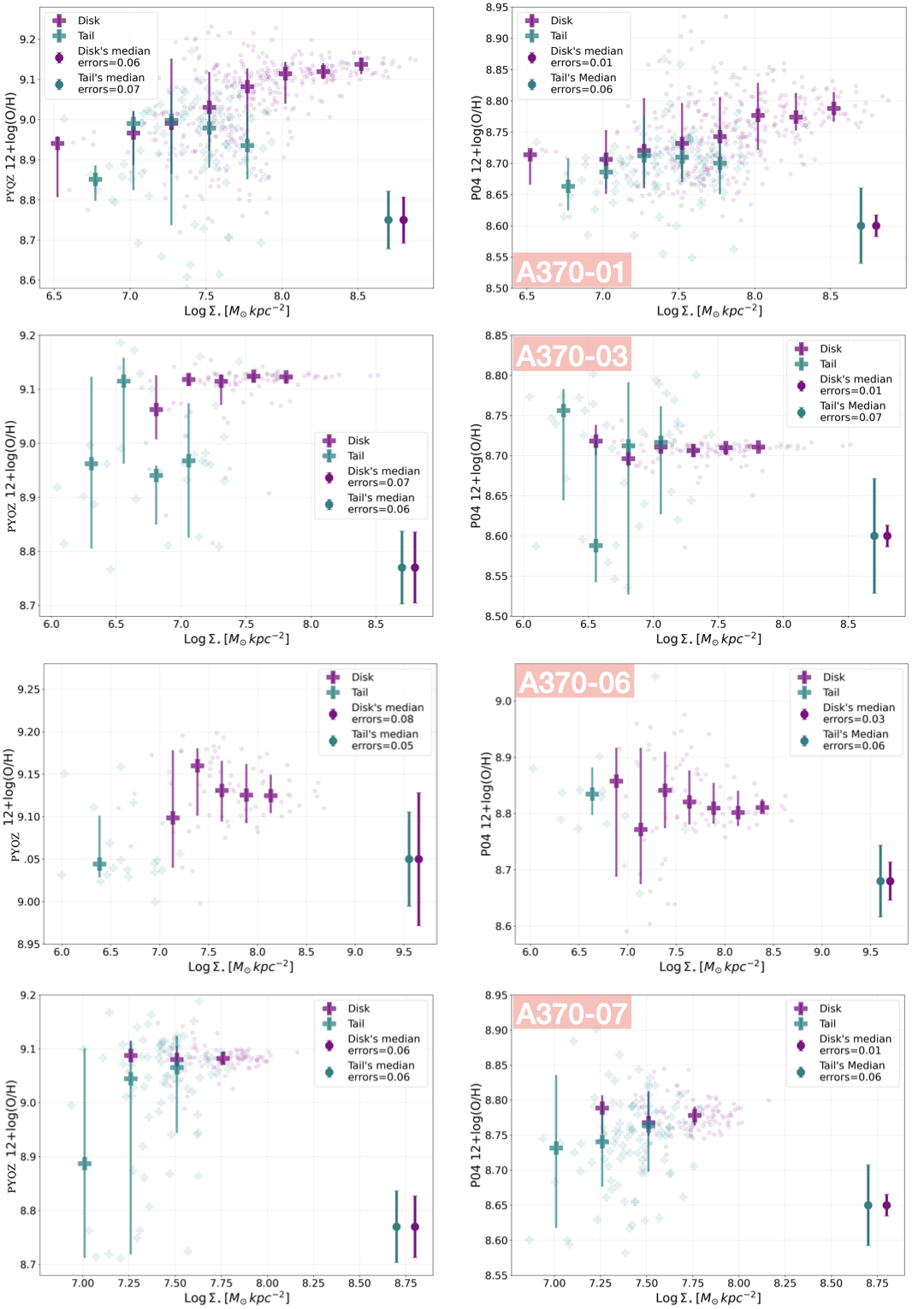}
    \caption{(continued)}
    \label{fig:MZR_TOT1}
\end{figure*}
% \clearpage

\setcounter{figure}{11}
\begin{figure*}[h!]
    % \figurenum{12}
    \centering
    \includegraphics[width=\linewidth,height=0.94\textheight,keepaspectratio]{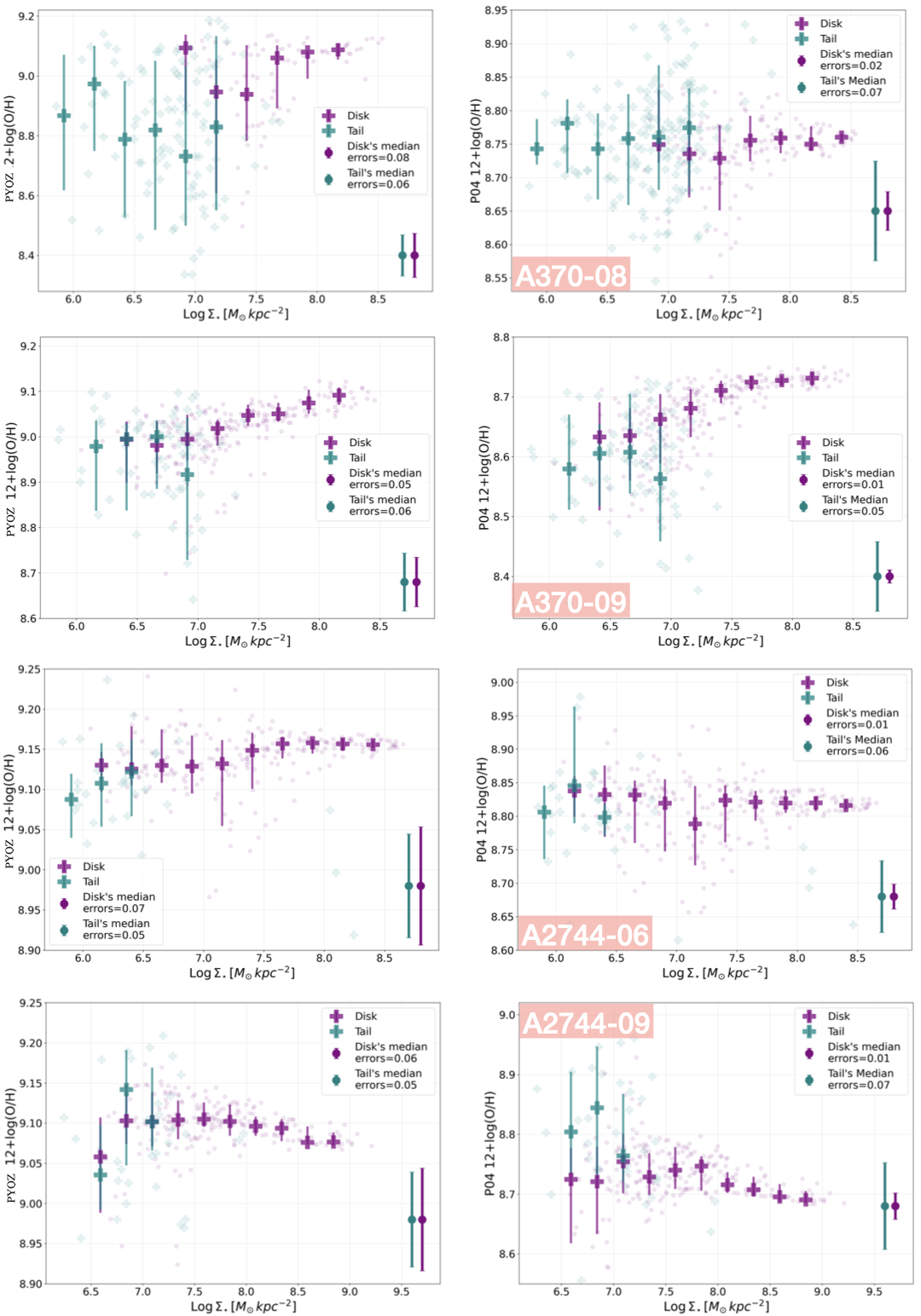}
    \caption{The rMZR of 8 RPS galaxies in A370 and A2744. The disk and tail spaxels with their accompanying metallicity uncertainty are color-coded in faded purple and teal, respectively. The bold plus markers provide the median metallicity values in stellar-mass surface density bins with $0.25$ dex width. Also, the median value errorbars represent the $1\sigma$ percentile deviation from the median value in each bin.}
    \label{fig:MZR_TOT}
\end{figure*}

% \clearpage
%\begin{figure*}
 %   \centering
  %  \includegraphics[width=\linewidth,keepaspectratio]{A370_01_MZR_Disk_Tail.png}
   % \caption{A370-01 rMZR using \textsc{pyqz} and P04 methods in the left and right plots. The disk and tail spaxels with their accompanying metallicity uncertainty are color-coded in purple and teal, respectively. The bold plus markers provide the median metallicity values in stellar mass bins with $0.25$ dex width. The median value errorbars represent the $1\sigma$ percentile deviation from the median value in each bin. }
    %\label{fig:A370_01 MZR}
%\end{figure*}
\subsubsection{Global Mass-Metallicity Relation}\label{subsec: MZR}

\begin{figure*}[h!]
    % \figurenum{13}
    \centering
    \includegraphics[width=0.95\textwidth]{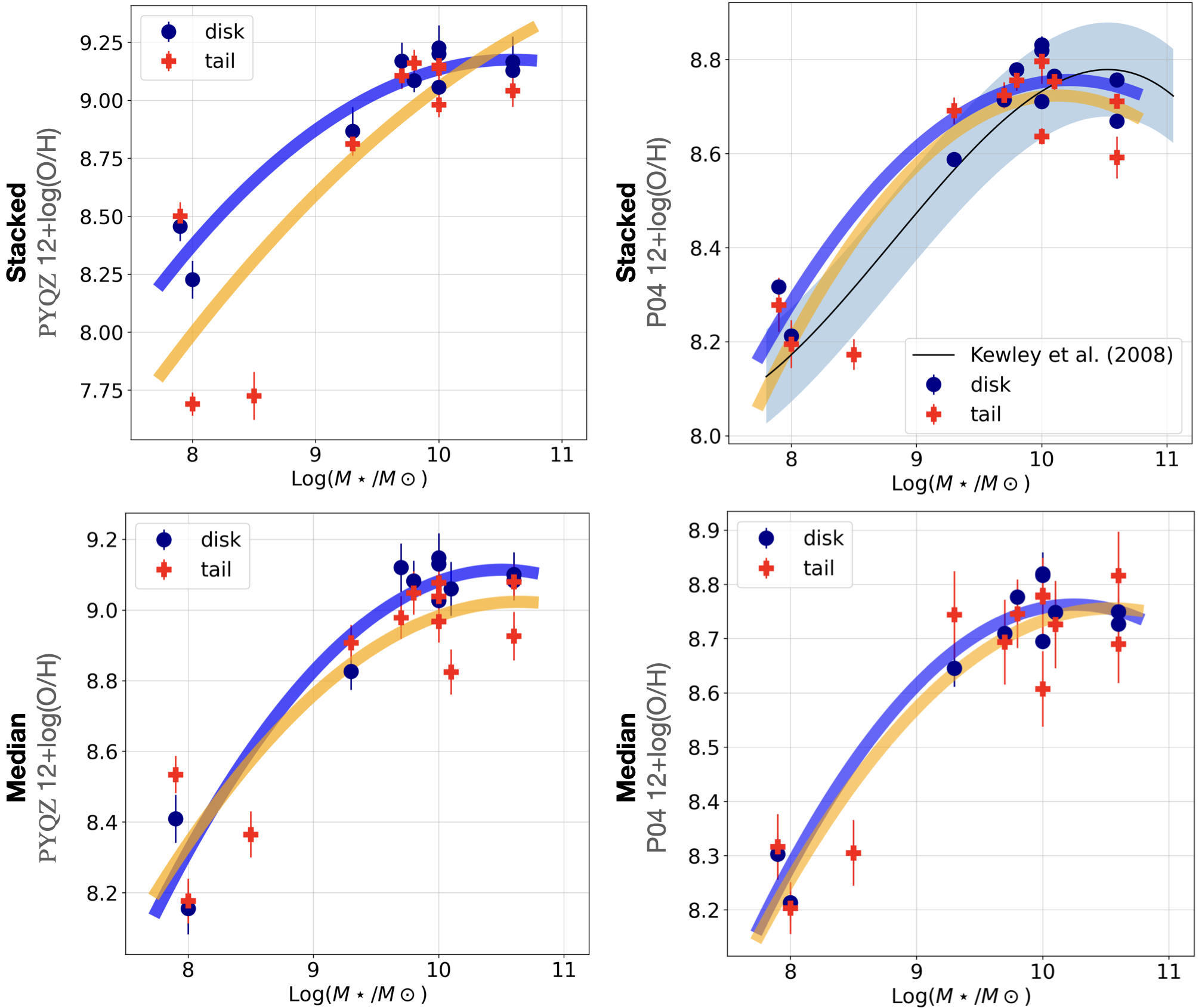}
    \caption{Measured global MZR of all galaxies in the sample from stacked spectra (top plots) and assigning global median metallicity values (bottom plots) of tails (red with orange fit) and disks (dark blue with blue fit) via second-order polynomial fit. The left and right panels represent \textsc{pyqz} and P04 MZR, respectively. In both panels, errorbars represent the metallicity uncertainty as discussed before. Additionally, due to the lack of usable individual spaxels in A2744-01, there is no median metallicity value available for the disk of this galaxy. In the right panel, Stacked and Median, the tail's metallicity of A2744-10 (i.e. the one with a steep positive gradient Fig.\ref{fig:Metgrad}) is higher than the disk's by $0.11$dex, exceptionally. The solid black line represents the \cite{2008ApJ...681.1183K}'s MZR fit with 0.1dex uncertainty illustrated with the blue shaded area.} 
    \label{fig: global MZR}
\end{figure*}

The top panel of Fig.\ref{fig: global MZR} shows the MZR (2nd order polynomial fit) that we derive for the stacked spectra of individual galaxies. For two reasons, disks and tails are investigated independently. Firstly, it is crucial to ascertain whether tails and disks exhibit a consistent mass-metallicity relation (MZR). Secondly, we can investigate the variation in global metallicity between the disk and the tail concerning parent galaxy's disk stellar mass and metallicity.

Considering Fig.\ref{fig: global MZR}'s top panel, for 
\textsc{pyqz} and P04, both disks and tails display a clear mass-metallicity relation. They notably follow a similar path with a positive slope (which is consistent with previous studies such as \citealt{2008A&A...488..463M, 2019MNRAS.484.3042S, 2020A&A...636A..84N}), and a flattening at high masses, in each metallicity diagnostic. Although \cite{2020ApJ...895..106F}, the sole analogous study available, utilized similar methodologies to those employed in this paper,  a direct comparison of our results is unfeasible due to the use of distinct apertures. 
We contrast our findings with the MZR proposed by \cite{2008ApJ...681.1183K}, as shown in the top right panel of Fig.\ref{fig: global MZR}. \cite{2008ApJ...681.1183K} employed a sample of 27,730 star-forming galaxies at z$<$0.1 obtained from the SDSS DR4 catalog (thus fibre-integrated spectra) using the \cite{1955ApJ...121..161S} IMF and the P04 metallicity calibration in their analysis. Converting the \cite{1955ApJ...121..161S} IMF to the \cite{2003PASP..115..763C} IMF, our findings are quite consistent with their relation within the reported uncertainty of 0.1 dex.\\

Furthermore, the global metallicity values of the disks and tails of individual galaxies exhibit a relative similarity, with the tail metallicities generally being slightly lower than those of their parent galaxy's disk, which is consistent with previous studies at low redshifts of individual galaxies by \cite{2016MNRAS.455.2028F}, \cite{2017ApJ...846...27G}, and the three galaxies by \cite{2021ApJ...923...28F}. Nevertheless, some exceptions exist, such as the galaxies A370-07 and A2744-03 in \textsc{pyqz} and a galaxy, A2744-10, in the P04 diagnostic, where the tail global metallicities are up to $\sim 0.1 dex$ higher than their corresponding disk values. Stacking spaxel spectra can result in biased derived properties if highly luminous spaxels with higher metallicity dominate the combined spectrum, similar to biases caused by stacking spectra with different star formation histories. %Therefore, 
By analyzing the median metallicity, we are taking into account the distribution of metallicities within the sample and are less susceptible to these biases. \\

 In the bottom panel of Fig.\ref{fig: global MZR} we present the MZR using the median metallicity value for the disk and tail regions, with the inclusion of only those spaxels that meet both the BPT diagram and SNR criteria (see Sec.\ref{subsec: spectral analysis}). Also in this case, the global metallicities of the disks and tails within individual galaxies demonstrate a degree of similarity. However, it is noteworthy that the metallicities of the tails are generally slightly lower than those of the disks in their respective parent galaxies, with the exception of three cases, namely A2744-03 in the \textsc{pyqz}, and A2744-09 and A2744-10 in the P04 metallicity diagnostic. Therefore, the overall trends remain consistent with those observed in Figure \ref{fig: global MZR}'s upper panel, where the MZR of the tail galaxies follows that of the disks, though with slightly but systematically lower metallicities at masses above $10^{9.3}{\text{M}_\odot}$. 

\begin{table*}[h!]
    \centering
    \resizebox{\textwidth}{!}{
\begin{tabular}{lcc|rrrr|rrrr}
\hline 
\multicolumn{3}{c}{ } &\multicolumn{4}{c}{P04} & \multicolumn{4}{c}{\textsc{pyqz}} \\
\hline 
 ID     & $\log(M_\star)$ &  z   &  Stacked Disk &   Stacked Tail  &  Median Disk &   Median Tail &  Stacked Disk &   Stacked Tail  &  Median Disk &   Median Tail  \\
\hline
 A370-01  &         10.6& 0.3738 &          8.76$\pm$0.001 &          8.71$\pm$0.01 &
 8.75$\pm$0.03  &         
 8.69$\pm$0.07  &                
 9.13$\pm$0.07 &          
 9.04$\pm$0.07 &   
 9.08$\pm$0.06 &         
 8.93$\pm$0.07
 
 \\
 
 A370-03  &          9.7& 0.3580
 &          8.71$\pm$0.003 &          
 8.72$\pm$0.03 &     
 8.71$\pm$0.02 &         
 8.69$\pm$ 0.08 &                    
 9.17$\pm$0.08  &         
 9.11$\pm$0.06 &    
 9.12$\pm$0.07  &         
 8.98$\pm$0.06\\
 
 A370-06  &         10.0& 0.3593 &          8.83$\pm$0.02 &          
 8.80$\pm$0.03 &       
 8.82$\pm$0.04 &         
 8.78$\pm$0.07&         
 9.23$\pm$0.10 &          
 9.15$\pm$0.09 &
 9.13$\pm$0.09  &       
 --           \\
 
 A370-07  &          9.8 & 0.3803&          8.78$\pm$0.01  &          
 8.76$\pm$0.02 &       
 8.78$\pm$0.02 &         
 8.74$\pm$0.06& 
 9.09$\pm$0.05 &          
 9.16$\pm$0.06 &  
 9.08$\pm$0.06 &         
 9.05$\pm$0.06\\
 
 A370-08  &         10.1& 0.3872 &          8.76$\pm$0.01 &          
 8.76$\pm$0.02 &      
 8.75$\pm$0.04 &         
 8.73$\pm$0.08&         
 --     &        
 --     &     
 9.06$\pm$0.08  &         
 8.82$\pm$0.06\\
 
 A370-09  &         10.0 & 0.3454 &          8.71$\pm$0.01 &          
 8.63$\pm$0.02 &      
 8.69$\pm$0.01 &         
 8.61$\pm$0.07&                  
 9.06$\pm$ 0.06 &          
 8.98$\pm$0.05 &  
 9.03$\pm$0.05 &         
 8.97$\pm$0.06\\
 
 A2744-01 &          8.5& 0.2919 &          -- &          
 8.18$\pm$0.03 & 
 --             &        
 8.31$\pm$0.06&        
 -- &          
 7.73$\pm$0.10 &  
 --     &         
 8.37$\pm$0.07\\
 
 A2744-03 &          7.9& 0.3026&          8.31$\pm$0.02 &         
 8.28$\pm$0.05 &         
 8.30$\pm$0.04 &         
 8.32$\pm$0.06&               
 8.46$\pm$0.06 &          
 8.50$\pm$0.06 &     
 8.41$\pm$0.07  &         
 8.54$\pm$0.05 \\
 
 A2744-04 &          8.0& 0.2941   &          8.22$\pm$0.03 &          
 8.20$\pm$0.05 &      
 8.21$\pm$0.04 &         
 8.20$\pm$0.05&              
 8.23$\pm$0.08  &          
 7.69$\pm$0.05  &        
 8.16$\pm$0.07 &         
 8.18$\pm$0.06 \\
 
 A2744-06 &         10.0 &0.2933  &          8.80$\pm$0.002  &          
 8.80$\pm$0.05 &       
 8.82$\pm$ 0.03  &         
 8.78$\pm$0.07&               
 9.20$\pm$0.08 &          
 9.14$\pm$0.06 &  
 9.15$\pm$0.07 &         
 9.08$\pm$0.05\\
 
 A2744-09 &         10.6& 0.2956 &          8.67$\pm$0.001 &          
 8.60$\pm$0.05&         
 8.73$\pm$0.04 &         
 8.82$\pm$0.08&              
 9.17$\pm$0.11 &        
 --     &               
 9.10$\pm$0.06 &         
 9.08$\pm$0.05\\
 
 A2744-10 &          9.3 & 0.2958&          8.59$\pm$0.001 &          
 8.69$\pm$0.03&         
 8.65$\pm$0.03 &         
 8.75$\pm$ 0.08       &               
 8.87$\pm$0.10  &          
 8.81$\pm$0.05 &     
 8.80$\pm$0.05 & 
 -- \\
\hline
\end{tabular}
}

    \caption{The P04 and \textsc{pyqz} stacked and median metallicity values of individual galaxies. The $M_\star$ values are in units of $M_\odot$. The uncertainties in the stacked spectra metallicity are determined according to the method outlined in Sec.\ref{sec: metallicity measurements}, and the median metallicity errors are obtained by averaging the uncertainties of spaxels within disks and tails.}
    \label{tab:P04 stack+med table+z+M}

\end{table*}

% \vspace{0.1cm}

Extraplanar tails of ionized stripped gas, extending up to several tens of kiloparsecs beyond the stellar disk, are often observed in ram-pressure stripped (RPS) galaxies in low redshift clusters. Recent studies have identified similar tails also at high redshift and we here present the first analysis of the chemical composition of such tails beyond the local universe. At first, we examine the resolved distribution of ionized gas metallicity of RPS galaxies in the Abell 2744 (z=0.308) and Abell 370 (z=0.375) clusters, the two most nearby clusters with the longest exposure time observed as part of the MUSE-GTO program. We investigate spatially-resolved and global metallicities in galactic disks and stripped tails, utilizing both a theoretical calibration through a photoionization model and an empirical calibration. The metallicity gradients and the spatially resolved mass-metallicity relations indicate that the metallicity in the tails reaches up to $\sim 0.6$dex lower values than anywhere in the parent disks, with a few exceptions. Both disks and tails follow a global mass-metallicity relation, though the tail metallicity is systematically lower than the one of the corresponding disk by up to $\sim 0.2$ dex. These findings demonstrate that additional processes are at play in the tails, and are consistent with a scenario of progressive dilution of metallicity along the tails due to the mixing of intracluster medium and interstellar gas, in accord with previous low-z results. Additionally, we examined all RPS in the other 10 clusters, 0.3<z<0.55, observed via the MUSE-GTO program alongside a control sample consisting of field and cluster galaxies. Through an examination of their disk chemical compositions, we unveiled unique characteristics exhibited by RPS galaxies, providing valuable insights into the interaction between the ISM and the ICM during ram-pressure stripping and the environmental impact on galaxy evolution at intermediate redshifts.

\section{Discussion}\label{sec: Discussion}

At low redshift, the gas metallicity of $\rm H\alpha$-bright clumps in RPS tails progressively decreases with increasing distance from the galaxy disk. This result was found for four strongly RPS galaxies at $z \sim 0.05$ by \cite{2021ApJ...922L...6F} and  \cite{2019MNRAS.485.1157B}. The metallicity gradients of high-mass RPS galaxies ($\log(M_\star/M_{\odot})>$9.3) in clusters at z=0.3-0.4 presented in this work generally show qualitatively similar trends to those observed in low redshift RPS massive galaxies (with one notable exception, A2744-09).

\cite{2021ApJ...922L...6F} presented three possible scenarios to explain the observed lower metallicity values in the tails of their three galaxies. The first scenario proposes that the progressively lower-metallicity gas observed further out in the tail might have been removed from the disk at large galactocentric radii, where typically is less metal-enriched, following the inside-out formation of disk galaxies. The authors dismissed this scenario as unrealistic because the stripped tails maintain their radius or get wider with distance from the disk, therefore the observed ionized gas cannot be stripped from well beyond the stellar disk. The same reasoning applies also to our sample, ruling out the first scenario.

The second scenario suggests that the observed metallicity values in the tails are due to the contamination of an additional ionization source that might alter the observed line ratios and metallicity measurements.
Scenario two was also disregarded by \cite{2021ApJ...922L...6F} because their analysis was conducted ensuring that only star-forming clumps were considered, based on the BPT diagram, though the authors noted that they could not exclude the possibility that exotic processes might produce artifacts in the metallicity measurement.
Regarding the second scenario for our sample, we note that emission which is classified as "composite" from the BPT diagram might be affected by ionization sources other than star formation.
However, the second scenario is unlikely also in our case, because a strong declining trend is observed in the tails also when considering only star-forming spaxels (e.g. A370-03, A370-08, and others).
It is worth noting that in their study, \cite{2021ApJ...922L...6F} incorporated an additional [OI]-[OIII] \cite{2001ApJ...556..121K} BPT demarcation line to differentiate star-forming regions. We tried applying these criteria also to our entire sample, finding that 87 percent of the mutually selected regions are considered star-forming based on our [OIII]-[NII] \cite{2001ApJ...556..121K} analysis. This indicates that our results are fully compatible with the findings of \cite{2021ApJ...922L...6F}, even when considering additional line ratios. Furthermore, in the [SII]-[OIII] BPT diagram \citep{2001ApJ...556..121K}, renowned for its sensitivity to detecting shocks, it is noteworthy that over 93\% of spaxels identified as star-forming or composite in this work reside below the \cite{2001ApJ...556..121K} [SII]-[OIII] BPT demarcation line. As an additional examination, we adopted the methodology proposed by \cite{2019D'Agostino} to investigate footage of shocks in the correlation between [OI]$\lambda$6300/H$\alpha$, [SII]$\lambda$6716,6730/H$\alpha$, [NII]$\lambda$6583/H$\alpha$, and velocity dispersion. Notably, our analysis revealed the absence of a robust correlation, thereby dismissing shock processes as a substantial source of contamination in our sample.

The third scenario postulated by \cite{2021ApJ...922L...6F} concludes that the low metallicity values in the tails are due to the mixing of the stripped ISM with the metal-poor ICM. Considering this 
the most plausible explanation, the authors provided a rough estimation of the mixing impact on the final gas metallicity. Interpreting our trends with the same arguments above, we conclude that the most likely explanation for our results is that the mixing of ICM and stripped ISM is efficient also at these redshifts and appears to be the underlying cause of the observed metallicity trends in the tails of distant cluster galaxies.
Moreover, in recent X-ray investigations \citep[][]{2019ApJ...887..155P,2021ApJ...911..144C,2022Bartolini}, it has been suggested that the pronounced X-ray emission observed in the tail is predominantly attributed to the ISM-ICM mixing, supporting the scenario where the stripped ISM undergoes heating due to its interaction with the ICM. These findings are also supported by numerical simulations studies \citep[e.g., ][]{2011Tonnesen,2012Tonnesen}, which have reported a wide distribution of electron density and temperature across both the tail and the disk. Therefore, in response to the heating observed in X-ray studies, we utilized realistic electron temperature values for HII regions to model the oxygen abundances in both disks and tails. Our results not only corroborate this heating phenomenon but also reinforce the concept of elevated temperatures within the tail. Specifically, our results demonstrate consistency with the trends highlighted in Sec.\ref{sec: results}, where similar findings emerge only when tail temperatures surpass those of their respective disks.\\

Regarding the ISM-ICM mixing scenario, further detailed studies might be helpful to shed light on ICM enrichment and its current metallicity in different regions since the enrichment of the tail involves a multitude of processes beyond ISM-ICM mixing. It is imperative to delve into a deeper understanding of various aspects, including the stripping stages of a galaxy, star formation within the tail, density and temperature distribution, and a comprehensive knowledge of ICM characteristics across different regions within a cluster.

\section{Summary and Conclusion}\label{sec: summary}

In this paper, we have studied the chemical evolution of galaxies at intermediate redshift that are undergoing ram-pressure stripping, investigating how intrinsic properties of galaxies, such as their stellar mass, influence the metallicity of the ISM and stripped-out gas in these entities. In order to do that, we used the data collected by MUSE via the MUSE GTO program and prior detection of RPS candidates in Abell370 and Abell2744 clusters provided by \cite{2022ApJ...925....4M}. \\

We focused on the gas content of 12 RPS galaxies whose
%with an emphasis on their chemical compositions. More specifically, their
metallicities were studied with theoretical and empirical methods. Two methods were adopted: a photoionization model, \textsc{pyqz}, and an empirical calibration, P04. We explored the metallicity gradient along the projected disk and tail of individual galaxies considering both star-forming and composite regions. Moreover, we studied spatially-resolved MZR of the disks and tails separately of those galaxies that contain meaningful stellar-continuum information. Also, the global MZR was studied via two approaches, stacking the spectrum of disks and tails and obtaining the median value of their metallicities. The main results of this study are listed below.\\

\begin{enumerate}[label=\roman*.]

\item P04 and \textsc{pyqz} generally agree on metallicity trends, except 
in very few cases. Moreover, The results from integrated spectra and median metallicity values are qualitatively consistent, regardless of using either the P04 or \textsc{pyqz} diagnostic.\\

\item  
%Note, 
In principle, composite emission is prone to inaccuracy due to other ionization sources such as AGN, LINER, shocks, or high turbulence. This issue is even more complex in RPS tails, where composite emission may arise from ISM-ICM mixing. The presence of numerous composite spaxels in some of the disks suggests either the effect of mixing on the disk or the misidentification of extraplanar spaxels as a disk.
The metallicity gradients of SF and composite spaxels show little systematic difference, as they usually follow a similar path. Although composite spaxels can be the most deviant in some disks, especially at low metallicities, they do not significantly alter the negative profiles or the comparison of metallicity measurement methods. \\

\item Most disk profiles are negative or flat. Tail profiles are usually negative, with varying slopes from the disk, likely due to projection effects. Also, in most cases, the majority of tail metallicities are lower than the disks. As a result, the overall gradient (disk + tail) is negative or nearly flat, with two exceptions. The present finding is consistent with the only existing study of ram-pressure stripped galaxies, which is at low redshift, \cite{2021ApJ...922L...6F}.\\

\item The vast majority of galaxies, which have a $\log(M_\star/M_{\odot})$ greater than or equal to 9.7, show tails with metallicities lower than those anywhere in their disks. %Also, 
Instead, the profiles across the disk and tail of the two lowest masses, approximately $\log(M_\star/M_{\odot})$ $\sim$ 8,  (with a high degree of uncertainty and only a few available points) show a trend consistent with being flat.\\

 \item The flat metallicity gradients of low stellar mass galaxies (log M $< 10^9 M_{\odot}$)
may be consistent with the mixing scenario too, considering that in low-mass galaxies the (low) metallicity of the interstellar medium can approach the value of the ICM metallicity. However, given the closeness of the disk angular size of these galaxies to the resolution limit of our observations,
it is hard to assess the meaning of this finding. 

\item The spatially-resolved relation between metallicity and surface stellar-mass density (rMZR) of the disks are generally flat or positive, with only one galaxy showing a slight negative trend. Tail spaxels, on the other hand, consistently display lower surface mass densities compared to their disk counterparts, regardless of the parent galaxy's stellar mass values (Tab.\ref{tab:P04 stack+med table+z+M}). The presence of tail spaxels with substantially lower metallicities than any disk spaxel with a similar surface mass density is 
evident from our data. This observation further reinforces the idea that different formation or enrichment processes are at work in the tails compared to the disk. \\

\item The global mass-metallicity relations of tails and disks in both \textsc{pyqz} and P04 exhibit the same trends. They share a similar positive slope, as observed in previous studies. Additionally, the global metallicity values of the tails and disks within each galaxy are relatively alike, but tail metallicities are generally slightly lower than those of the disk (except for a few galaxies in which the trend is inverted), which is consistent with previous studies by \cite{2020ApJ...895..106F}, \cite{2017ApJ...846...27G}, and \cite{2021ApJ...922L...6F}. The most likely explanation for such trends is the mixing of the enriched gas stripped from quite massive galaxies with the lower metallicity ICM.

\end{enumerate}

In conclusion, the observations presented in this study suggest that the metallicity distribution in the tails of galaxies is shaped both by the metallicity of the parent galaxy and by the mixing with the metal-poor intracluster medium as a result of ram-pressure stripping. This finding is supported by the observed steep negative profiles in massive (metal-rich) galaxies and lower global metallicity values of tails compared to their parent disks. 
\\

We thank the referee for helpful comments that improved the paper. This project, funded by the European Research Council (ERC) under the European Union's Horizon 2020 research and innovation programme (grant agreement No. 833824), is based on observations gathered at the European Organization for Astronomical Research in the Southern Hemisphere, under the ESO programme 196.B-0578. This work used products from HFF-DeepSpace, funded by the National Science Foundation and Space Telescope Science Institute (operated by the Association of Universities for Research in Astronomy, Inc., under NASA contract NAS5-26555).

%--------------------------------------------------------------------

%-------------------------------------- Two column figure (place early!)

% WARNING
%-------------------------------------------------------------------
% Please note that we have included the references to the file aa.dem in
% order to compile it, but we ask you to:
%
% - use BibTeX with the regular commands:
\clearpage
\newpage
  \bibliographystyle{aa} % style aa.bst
  \bibliography{ref.bib} % your references Yourfile.bib
%
% - join the .bib files when you upload your source files
%-------------------------------------------------------------------

\end{document}